\documentclass[a4paper,fleqn,usenatbib]{mnras}
\usepackage{newtxtext,newtxmath}
\usepackage[T1]{fontenc}
\usepackage{ae,aecompl}

\usepackage{graphicx}	
\usepackage{amsmath}	

\newcommand{\msun}{\mbox{$\,{\rm M}_\odot$}}
\newcommand{\oversim}[2]{\protect{\mbox{\lower0.5ex\vbox{%
  \baselineskip=0pt\lineskip=0.2ex
  \ialign{$\mathsurround=0pt #1\hfil##\hfil$\crcr#2\crcr\sim\crcr}}}}}
\newcommand{\simgreat}{\mbox{$\,\mathrel{\mathpalette\oversim>}\,$}} 
\newcommand{\simless} {\mbox{$\,\mathrel{\mathpalette\oversim<}\,$}} 

\title[Formation of UCDs]{Do ultra compact dwarf galaxies form monolithically or as merged star cluster complexes?}

\author[Mahani et al.]
{Hamidreza Mahani$^{1}$,
 Akram Hasani Zonoozi$^{1,2}$\thanks{E-mail:  \mbox{a.hasani@iasbs.ac.ir} }, Hosein Haghi$^1$\thanks{E-mail:  \mbox{haghi@iasbs.ac.ir} }, Tereza Je{\v r}{\'a}bkov{\'a}$^{4}$\thanks{ESA fellow },
 \newauthor
 Pavel Kroupa$^{2,3}$, Steffen Mieske$^5$ \\
$^{1}$Department of Physics, Institute for Advanced Studies in Basic Sciences (IASBS), PO Box 11365-9161, Zanjan, Iran\\
$^{2}$Helmholtz-Institut f\"ur Strahlen-und Kernphysik (HISKP), Universit\"at Bonn, Nussallee 14-16, D-53115 Bonn, Germany\\
$^{3}$Charles University in Prague, Faculty of Mathematics and Physics, Astronomical Institute, V Hole\v{s}ovi\v{c}k\'ach 2, CZ-180 00 \\
Praha 8, Czech Republic\\
$^4$European Space Research \& Technology Centre ,Keplerlaan 1, Postbus 299, 2200 AG Noordwijk, Netherlands\\
$^5$European Southern Observatory, Alonso de Córdova 3107, Vitacura, Casilla 19001, Santiago, Chile\\}

\date{Accepted XXX. Received YYY; in original form ZZZ}

\pubyear{2019}

\begin{document}
\label{firstpage}
\pagerange{\pageref{firstpage}--\pageref{lastpage}}
\maketitle

\begin{abstract}
Some ultra-compact dwarf galaxies (UCDs) have elevated observed dynamical V-band  mass-to-light ($M/L_{\rm V}$) ratios with respect to what is expected from their stellar populations assuming a canonical initial mass function (IMF). Observations have also revealed the presence of a compact dark object in the centers of several UCDs, having a mass of a few to 15\% of the present-day stellar mass of the UCD. This central mass concentration has typically been interpreted as a super-massive black hole, but can in principle also be a sub-cluster of stellar remnants. We explore the following two formation scenarios of UCDs, i) monolithic collapse and ii) mergers of star clusters in cluster complexes as are observed in massively star-bursting regions. We explore the physical properties of the UCDs at different evolutionary stages assuming different initial stellar masses of the UCDs and the IMF being either universal or changing systematically with metallicity and density according to the Integrated Galactic IMF (IGIMF) theory. While the observed elevated $M/L_{\rm V}$ ratios of the UCDs cannot be reproduced if the IMF is invariant and universal, the empirically derived IMF which varies systematically with density and metallicity shows agreement with the observations. Incorporating the UCD-mass-dependent retention fraction of dark remnants improves this agreement. In addition we apply the results of N-body simulations to young UCDs and show that the same initial conditions describing the observed $M/L_{\rm V}$ ratios  reproduce the observed relation between the half-mass radii and the present-day masses of the UCDs. The findings thus suggest that the majority of UCDs that have elevated $M/L_{\rm V}$ ratios could have formed monolithically with significant remnant-mass components that are centrally concentrated, while those with small $M/L_V$ values may be merged star-cluster complexes.
\end{abstract}

\begin{keywords}
stars: luminosity function, mass function -- Galaxy: globular clusters -- Galaxy: open clusters and associations -- galaxies: formation -- galaxies: interactions -- galaxies: starburst
\end{keywords}


\section{Introduction}

Ultra-compact dwarf galaxies (UCDs) are typically found in group and cluster environments and have present-day dynamical masses in the range of $10^{6} \simless M/\msun \simless 10^{8} \msun $ and half-mass radii  $r_{\rm h} < 100\,$pc \citep{Hilker99, Drinkwater2000}. They are characterised by average V-band luminosities of $ -14.5 \simless  M_{\rm V} \simless -10.5 $ mag and central velocity dispersions of $ 20 \simless \sigma/({\rm km/s}) \simless 50$ \citep{Evstigneeva07}. The mass-to-light ($M/L_{\rm V}$) ratios of many UCDs are larger than the $M/L_{\rm V}$ ratios of globular clusters (\cite{Mieske08, Mieske13}; see also \citealt{Voggel19}). Being rare, UCDs were discovered in galaxy clusters like Fornax and Virgo \citep{Hilker99}, but, they have also been found to be associated with galaxies in the field \citep{Evstigneeva07}.


Until now there is no consensus on whether  UCDs are the result of the evolution of composite stellar systems like galaxies, let alone whether they are formed with an unusual stellar initial mass function (IMF). The detection of a central excess mass (e.g. \citealt{Seth2014}) is often interpreted to be due to a super massive blacke hole (SMBH) and thus as  smoking gun evidence for the galaxy origin. Different formation scenarios have been proposed to explain the origin of UCDs:

\begin{enumerate}
\item \textbf{Remnant nuclei of galaxies:} UCDs maybe the remnant nuclei of dwarf elliptical galaxies which were tidally stripped \citep{Bekki01}. However, the observed number of these objects with masses larger than $ 10^7 \msun $ is two times larger than what is expected from the $\Lambda$CDM Millenium~II simulation \citep{Pfeffer14}, which is interpreted (assuming $\Lambda$CDM to be valid, \citealt{Kroupa_2012, Haslbauer_2020, Asencio_2020}) to mean that UCDs might be a mix between stripped nuclei and massive genuine star clusters \citep{Voggel19, Seth2014}.

 \item \textbf{Early cosmological fluctuations of dark matter:} UCDs may be the final stage of primordial compact galaxies formed in the early universe \citep{Drinkwater04}. However, the kinematics and the distribution of baryonic mass do not indicate the domination by non-baryonic dark matter halos at spatial scales of a few tens of pc, typical for UCDs \citep{Murray09, Strader13}. This was later  supported by a detailed study of one of the brightest UCDs in the Fornax cluster \citep{Frank11}. Also, for this scenario to be relevant, dark matter particles need to exist \citep{Kroupa15}.
 \item \textbf{Merging star clusters:} UCD-type objects may form from many merged star clusters \citep{Fellhauer02, Maraston04, Bruns11}. The simulations of merging "superclusters" by \citet{Kroupa98} predicted the existence of "spheroidal dwarf galaxies" (later found as UCDs). Such superclusters with radii of roughly 100~pc are observed in the tidal tail of the Tadpole galaxy (figs. 4 and 10 in \citealt{Kroupa15}), and as star cluster complexes in the inner region of the strongly interacting Antennae galaxies \citep{Kroupa98}. Observations such as these suggest that massively interacting gas-rich galaxies lead to extremely massive molecular cloud formation within which dense clusters of massive monolithically-formed embedded star-burst clusters form\footnote{``Monolithic'' formation is here referred to as the formation on a cloud-core free-fall time scale \citep{BK15, BK18}}.  Also, the simulations of colliding galaxies show that if two gas-rich galaxies interact strongly, gas clouds are pressurised and form stars in clusters with masses in the range $ 10^{7}-10^{8} \msun $ and with the size of a few hundred pc \citep{Elmegreen93, Bournaud_2008, Renaud_2015, Maji_2017}.

\item \textbf{Monolithic collapse:} Galaxies with larger star formation rates (SFRs) are observed to have more massive most-massive very young clusters which can be explained if the galaxy populates the embedded cluster mass function completely every $\delta t \approx 10\,$Myr with all stars forming in embedded clusters with masses larger than about $5\,M_\odot$ \citep{Weidner04b, R+13}.
Some UCDs may therefore be the most massive star clusters that can be formed in galaxies with  SFRs $\geq 100 M_{\odot}/yr$. They may be an extension of clusters such as the ONC, NGC3603, and R136 which are too young and too compact to have formed through the merging of sub-clusters \citep{KAH, BK14, BK15, BK18}. \cite{Murray09} develops a theoretical description of UCDs forming monolithically from collapsing cloud-cores finding that for cloud core masses $\simgreat 3\times 10^6\,M_\odot$ these become opaque for infrared radiation such that the IMF emerging therein becomes top-heavy.
 \end{enumerate}

In the Milky Way (MW), which presumably is a typical disk galaxy, the molecular clouds follow a mass--radius relation, $M_{\rm mcl}/M_\odot \approx 36.7\,\left(R_{\rm mcl}/{\rm pc} \right)^{2.2}$ \citep{MD+17}. The free-fall time, $t_{\rm ff}=\left( 3\,\pi / (32 \, G \, \rho_{\rm mcl}) \right)^{1/2}$ with the density $\rho_{\rm mcl} = 3\,M_{\rm mcl} / (4\, \pi\, R_{\rm mcl}^3)$ such that $ t_{\rm ff}/{\rm Myr} = 1.37 \, \left(M_{\rm mcl}/M_\odot \right)^{0.18}$. This implies that under MW conditions more massive clouds are likely to form more, and more massive, individual embedded clusters but that these take longer to fall towards each other the more massive the cloud is. The cloud is thus likely to disperse due to the ionising and wind energy produced by the massive stars in the embedded clusters before they can merge \citep{Vazquez-Semadeni+17}.  But in strongly interacting gas-rich galaxies the radii of molecular clouds are smaller for a given mass because the inter stellar medium is pressurised through the encounter, and this allows the massive embedded clusters to merge forming a more massive cluster from the cluster complex or supercluster in extreme cases. For example, the strongly-interacting Antennae galaxies are observed to be profusely forming massive superclusters \citep{Kroupa98, WhitmoreSchweizer95, Whitmore+99}, while post-encounter disk galaxies probably formed, during the encounter, less-massive versions of these which today are evident as ``faint-fuzzy'' star clusters \citep{Brodie_2002, Bruens+09}.
 In the Local Group of galaxies, monolithic formation has been the dominant process for the embedded, open and globular star clusters, as has been shown by the above-mentioned detailed modelling of the ONC, NGC3603 and R136 to be the case. Nevertheless, the massive and distended GC NGC~2419 \citep{BruensKroupa11}  and possibly $\omega$~Cen \citep{FK03}
bear signatures that they may have formed as merged cluster complexes.

It is thus of interest to investigate which role this merging process plays in contributing to the UCD population. While it is quite likely that all formation scenarios play a role in establishing the population of UCDs, the problems mentioned above with scenarios~(i) and~(ii) lead us to investigate here the last two mentioned scenarios, i.e.,the merging and monolithic collapse process.  Important empirical  constraints are the dynamical $M/L_{\rm V}$ ratio of all UCDs, some of which are found to be higher than expected for a simple stellar population assuming the canonical IMF \citep{Kroupa01b}, and the radius--mass relation of UCDs which departs from that of open and globular clusters (GCs) \citep{Dabringhausen08}. Assuming a universally-valid formulation of the stellar IMF as a function of the metallicity and density of the star-forming gas on the embedded-cluster (molecular cloud core) scale,  the key insight is that if a UCD forms from merged star clusters then the summed-IMF will be less top-heavy than the IMF of the equally-massive monolithically-formed UCD \citep{Kroupa03c, Kroupa13}. This leads to a dynamical $M/L_{\rm V}$ ratio which is smaller if a UCD formed through merging than via monolithical collapse, allowing a possible distinction of the formation process. The resulting radius of the UCD can then be used as an additional constraint on the formation process.

This paper is organised as follows: In Section 2 we describe the calculation method and the stellar population models that are constructed for UCDs based on the monolithic and merging scenarios. The main results are presented in Section 3.  Finally, Section 4 contains the discussion and conclusions.

\section{Method}

First the computation of the stellar population and the IMF in the different scenarios is introduced, then the stellar population synthesis (SPS) technique to study the formation and the evolution of UCDs is described.

\subsection{The stellar IMF}

The shape of the stellar IMF in an embedded star cluster (i.e. what comprises the full stellar population which forms in one molecular cloud core within $\simless 1\,$pc)
is a focal topic of investigation as it determines the high-mass stellar content and hence the dynamics of the cluster during and after its embedded phase.  Most studies of resolved stellar populations have shown the stellar IMF to be largely universal \citep{Kroupa01b, Kroupa13, Hopkins18}, with the invariant canonical IMF being well described  by a two-part power-law function. The number of stars with mass in the interval of $m$ (in units of $M_\odot$) and $m+dm$ is $dN=\xi(m)dm$, with      
\begin{equation}
  \xi(m \leq m_{\rm max})= k \begin{cases}
    2m^{-\alpha_{1}}, & \text{0.08\msun $ \leq $ m < 0.5\msun},\\
    m^{-\alpha_{2}}, & \text{0.5\msun $ \leq $ m < 1\msun},\\
    m^{-\alpha_{3}}, & \text{1\msun $ \leq $ m $ \leq $ $m_{\rm max}(M_{\rm ecl})$} \le m_{\rm max*},\\
  \end{cases}
  	\label{eq:imf}
\end{equation}
where, $ \alpha_{1}=1.3 $, $ \alpha_{2}=2.3 $ and $k$ is the normalization constant which depends on how many stars are formed in the embedded cluster, being determined through eq. \ref{eq:mmax} below. For the invariant canonical IMF, $ \alpha_{3}=\alpha_{2}=2.3 $ and  $m_{\rm max*}=150 M_\odot$ is the physical upper limit of stellar masses (e.g. \citealt{Koen06,MA+07,Banerjee+12}). The $m_{max}(M_{\rm ecl})$ relation is the mass of the  most massive star that can form in an embedded cluster with a total initial stellar mass of $M_{\rm ecl}$. Solving the following two equations, it is possible to determine both normalization constant, $k$ and $m_{\rm max}(M_{\rm ecl})$ \citep{Weidner06,Weidner10,Weidner13,OK18},

\begin{equation}
\begin{aligned}
    M_{\rm ecl}=\ \int_{0.08 \msun}^{m_{\rm max}} m \xi(m) dm, \\
    1=\ \int_{m_{\rm max}}^{m_{\rm max*}} \xi(m) dm,
	\label{eq:mmax}
\end{aligned}
\end{equation}
the first equation being the mass in stars with mass above the hydrogen-burning mass limit ($\approx 0.08\,M_\odot$) in the very young UCD (or embedded cluster) and the second equation constitutes the statement that there is one most massive star in the range $m_{\rm max}$ to $m_{\rm max*}$. Note that super-canonical stars ($m>m_{\rm max*}$) can arise due to binary mergers in massive young clusters \citep{Banerjee+12,OK18}.

On theoretical grounds, it is expected that the IMF ought to vary with the physical conditions.  Thus, it is expected to become more top-heavy (i.e., overabundant in high-mass stars) for high temperature and metal-poor star-forming regions as based on both the Jeans-mass instability argument \citep{Larson98, Murray09} and self-feedback-regulation arguments \citep{Adams96a, Adams96b, Kroupa13}. Also, under extremely dense conditions, proto-stellar cores are likely to merge before they can spawn a proto-star such that the mass-distribution of forming stars shifts to an over-abundance of massive stars \citep{Dib+07}.  Furthermore, \cite{Papadopoulos+11} suggest that in cosmic-ray dominated regions (e.g. in the vicinity of star-bursts) the cloud temperatures are raised significantly leading to a top-heavy IMF.

Evidence for a systematically varying IMF has emerged through the higher dynamical $M/L_{\rm V}$ ratios of some UCDs \citep{Dabringhausen09, Mieske08, MieskeKroupa08} and, independently, the larger incidence of low-mass X-ray binaries in some UCDs in the Virgo cluster \citep{Dabringhausen12} which together imply a consistent variation of the IMF. Interestingly, \cite{Phillipps_2013} have found that UCDs in the Fornax cluster were less likely to contain  low-mass X-ray binaries than GCs. Due to their extremely high birth binding energy,  UCDs that are born with a top-heavy IMF survive the residual gas expulsion process and stellar-evolutionary mass loss \citep{Dabringhausen10}. It remains unclear whether these  elevated mass-to-light ratios result from a systematically varying IMF that causes the average stellar mass to be higher than expected  (leaving many dark remnants) or whether it is due to the presence of SMBHs \citep{Seth2014}. Indeed the existence of SMBHs and top-heavy IMFs may be directly causally linked \citep{Kroupa20}.

Noteworthy is that the deficit of low-mass stars in GCs which have a low concentration is also explainable with the same IMF variation (\citealt{Marks12b}; Eq. \ref{eq:alpha3} below). A shallower mass function for metal-richer GCs can explain the declining optical and near-infrared dynamical mass-to-light ratios of M31 GCs with increasing metallicity \citep{Strader11}, in apparent contradiction with the observed higher than expected $M/L_V$ values of the most metal-rich UCDs.
This convergence of completely independent analysis methods and data suggests the resulting IMF dependency on metallicity and density as formulated by \cite{Marks12b} based largely on GC data to be useful as a hypotheses to be employed in further investigations. According to this formulation, the IMF becomes increasingly top-heavy with decreasing metallicity and increasing star-forming-gas density in the embedded-cluster-forming cloud core and has been shown to be consistent with the observed variations on galaxy-wide scales \citep{Yan+17}.

Corroborating this IMF variation, near the Galactic center an extremely flat IMF with $ \alpha_3 \approx 0.4 $ is obtained for stars formed in situ in a single star formation event about 6 Myr ago \citep{Bartko10}. The R136 star-burst cluster in the Large Magellanic Cloud must have had a slightly top-heavy IMF if the ejected massive stars are accounted for \citep{BK12}.  Indeed, the whole 30~Dor star-forming region has been found to have an excess of massive stars \citep{Schneider+18} and the low-metallicity Magellanic-Bridge cluster NGC~796 has also been observed to have a top-heavy IMF \citep{Kalari+18}. This empirical evidence is consistent with the formulation by \cite{Marks12b}.

The systematically varying stellar IMF with physical conditions of star formation regions (Eq. \ref{eq:alpha3} below) is also adopted to explain the unexpected observed correlation between $M/L_{\rm V}$ and metallicity of GCs in the Andromeda galaxy\citep{Strader11, Zonoozi16, Haghi17}. However, it should be mentioned that other possible solutions  to the derived metallicity--dynamical-mass-to-light ratio correlation of GCs are mass segregation \citep{shanahan_2015} and the preferential loss of low-mass stars in the stronger tidal fields experienced by metal-rich GCs \citep{Bianchini_2017, Baumgardt_2018}. Most recently, \cite{Baumgardt_2020} showed that accounting for alpha-abundance variations with metallicity improve the comparison between measured and predicted mass-to-light ratios of Milky Way GCs.

The dependence of $\alpha_3$ on both metallicity and density of the embedded-cluster can be described as \citep{Marks12b}

\begin{equation}
  \alpha_{3}= \begin{cases}
    2.3, & \text{x $ \leq $ -0.87},\\
    -0.41x + 1.94, & \text{x > -0.87},\\
  \end{cases}
\label{eq:alpha3}
\end{equation}
where, $ x=0.99$ log$_{10}(\rho_{\rm tot}/10^{6} \msun {\rm pc}^{-3})-0.14\,[Z/Z_\odot]$ which\footnote{Note that [Fe/H] in the original formulation by \cite{Marks12b} has been here replaced by $[Z/Z_\odot]$. This change is inconsequential for the models. Also, in Eq. \ref{eq:imf} $\alpha_1$ and $\alpha_2$ may depend on metallicity, as discussed in Sec. 6.6 in \cite{Yan_2020}}. implies  dense star clusters to have a top-heavy IMF with  $ \alpha_{3}<2.3 $.
The birth density of the embedded cluster is given by
the total gas and stellar density,
\begin{equation}
  \rho_{\rm tot}=\frac{3 M_{\rm tot}}{4\epsilon \pi r_{\rm h0}^3}.\\
\end{equation}
Its initial or birth half-mass radius has been derived by \citet{Marks12} by calculating the densest phase an observed sample of star clusters were allowed to have given the observed widest binary-star orbits in them\footnote{Note that for $M_{\rm ecl}=1 M_{\odot}$ the observed thickness of molecular cloud filaments \citep{Andre14} is obtained.},
\begin{equation}
  r_{\rm h0}/{\rm pc}= 0.1(M_{\rm ecl}/M_{\odot})^{0.13}.\\
\label{eq:rMrel}
\end{equation}
For a star formation efficiency $\epsilon$, the total (gas plus stellar) mass is $M_{\rm tot}= M_{\rm ecl}/\epsilon$. Assuming $\epsilon=1$, $M_{\rm tot}=M_{\rm ecl}$, being a reasonable approximation given that the binding energy of the born UCD is likely to surpass the feedback energy from its stars (e.g. fig.~3 in \citealt{Baumgardt+08}). This is strictly only true for an invariant canonical IMF but since the number of massive stars in the young UCD scales slightly superlinearly with UCD mass, the quadratic increase with UCD mass of the binding energy wins.

Therefore, an initially more massive embedded cluster has a more top-heavy IMF. Table~\ref{tab:alpha3} shows how $\alpha_3$ depends on $M_{\rm ecl}$ and on $Z$.

\begin{table}
 \begin{center}
 \caption{ Values of $\alpha_{3}$ for different initial masses and  metallicities for the variable IMF \citep{Marks12b} and for $\epsilon=1$.}
 \begin{tabular}{cccc}
 \hline
 \hline
 $\rm log_{10}\left(\frac{M_{\rm ecl}}{\msun}\right)$  & $ \alpha_3$ &$\alpha_3$ &$ \alpha_3$\\
   & $ (Z=0.02)$ &$ (Z=0.004)$ &$ (Z=0.0004)$
 \\
 \hline
  $5.0$  & $ 1.98$ &$ 1.94$ &$ 1.88$  \\

  $5.5$  & $1.85 $ &$ 1.81$ &$ 1.76$  \\

  $6.0$  & $ 1.73$ &$ 1.69$ &$1.63$  \\

  $6.5$  & $ 1.60$ &$ 1.56$ &$ 1.51$  \\

  $7.0$  & $ 1.48$ &$ 1.44$ &$ 1.38$  \\

  $7.5$  & $ 1.36$ &$ 1.32$ &$ 1.26$  \\

  $8.0$  & $ 1.23$ &$ 1.19$ &$ 1.14$  \\

  $8.5$  & $ 1.11$ &$ 1.07$ &$ 1.01$  \\

  $9.0$  & $ 0.99$ &$ 0.95$ &$ 0.89$  \\
 \hline
 \hline

 \end{tabular}
 \label{tab:alpha3}
 \end{center}
 \end{table}

 \begin{figure*}
\begin{center}
\includegraphics[width=150mm,height=120mm]{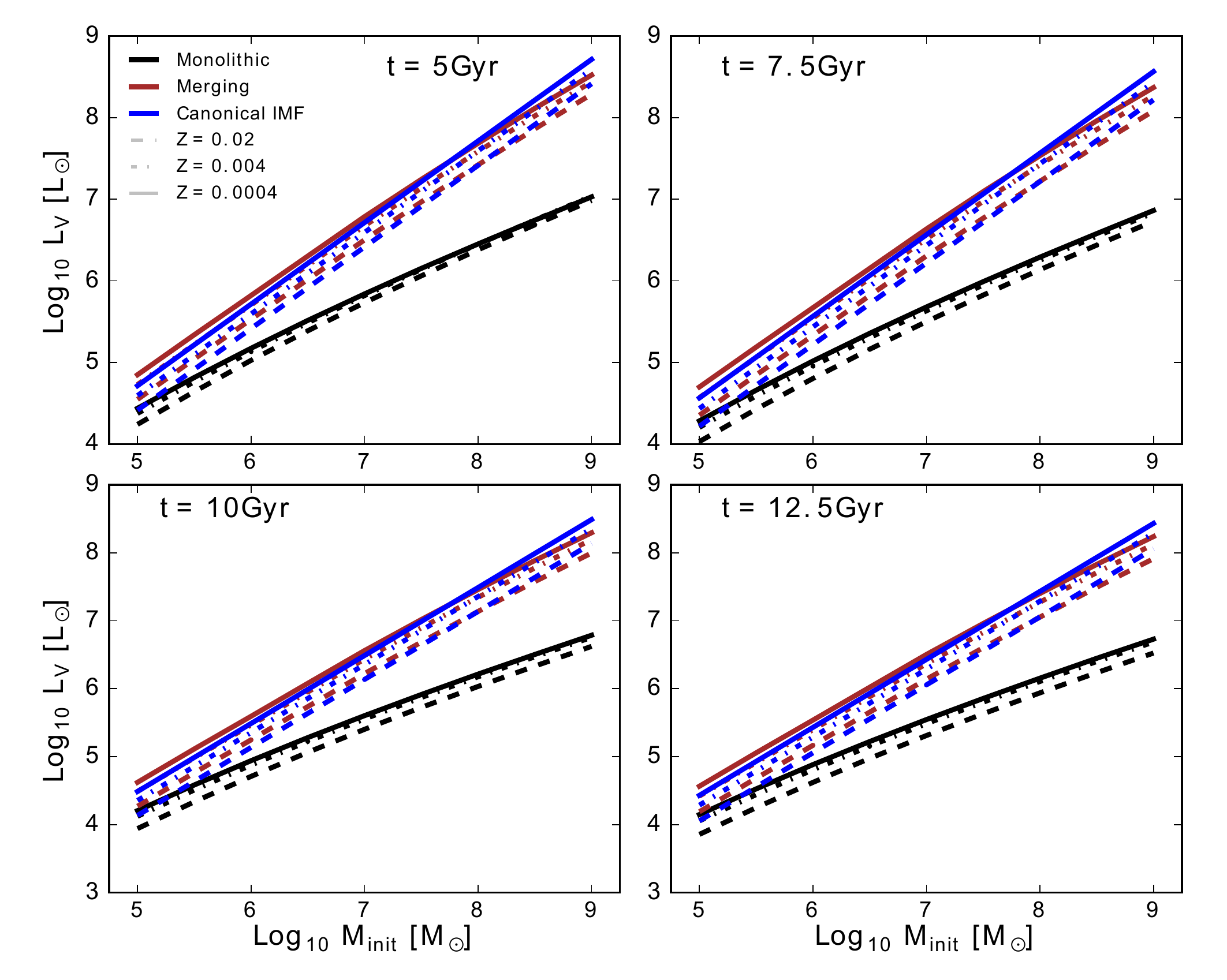}
\caption{The $V-$band luminosity of the UCD models as a function of their initial mass for different formation mechanisms and different values of metallicity as indicated in the key. The resulting $V-$band luminosity assuming the canonical IMF, monolithic collapse and the merging scenario are shown as blue, black and red lines, respectively. }
\label{fig:LV}
\end{center}
\end{figure*}

\subsection{The IMF of stars in UCDs formed monolithically}

In this case the IMF is equal to $\xi(m)$ (Eq.~\ref{eq:imf} and~\ref{eq:alpha3}) and the initial mass of all stars that formed in the UCD is $M_{\rm init}=M_{\rm ecl}$.

\subsection{The IMF of stars in UCDs formed as merged cluster complexes}

In order to study how the merging cluster-scenario affects the properties of UCDs, the mass function of the initial distribution of clusters within the cluster complex needs to be assumed. We use the empirically constrained power-law mass function of embedded clusters in galaxies (the ECMF, see bellow). It is possible, however, that in the case of UCDs, the embedded-cluster mass function might differ or not be fully populated. Thus our assumption serves as a certain borderline case — for example, if a UCD  formed from only two large clusters,  its final properties would be in-between the monolithically formed and the fully-populated ECMF cases.

In the merging scenario, UCDs form  through the merging  of many star clusters in cluster complexes. The notion is that each cluster in the complex forms monolithically according to the rules described above. The star clusters in the complex (referred to here as embedded clusters) with different masses encounter and merge with each other and  result in  a massive object within a time-scale of typically 100~Myr and with a larger radius \citep{Bruns11} and a composite IMF, $\xi_{\rm comp}$, that is calculated according to the IGIMF theory by adding the IMFs of all merged star clusters,
\begin{equation}
	\xi_{\rm comp}(m)=\int_{M_{\rm ecl}^{\rm min}}^{M_{\rm ecl}^{\rm max}} \xi(m \leq m_{\rm max}) \xi_{\rm ecl}(M_{\rm ecl})dM_{\rm ecl},
		\label{eq:mergeIMF}
\end{equation}
where, $ \xi(m \leq m_{\rm max}) $ is the IMF in each embedded cluster (Eq.~\ref{eq:imf} and~\ref{eq:alpha3}).
The embedded cluster mass function (ECMF), $\xi_{\rm ecl}(M_{\rm ecl})$, in the forming UCD is assumed to be a single power law function,
\begin{equation}
	\xi_{\rm ecl}(M_{\rm ecl})= k_{\rm ecl}M_{\rm ecl}^{-\beta} \; ,
		\label{eq:powerlaw}
\end{equation}
with the number of embedded clusters with stellar masses in the interval $M_{\rm ecl}$ and $M_{\rm ecl}+dM_{\rm ecl}$ being $dN_{\rm ecl}=\xi_{\rm ecl}(M_{\rm ecl}) dM_{\rm ecl}$.
For simplicity we assume that the power-law index of the embedded cluster mass function is a constant value of $\beta=2$, but note that the ECMF may become top-heavy at low metallicity \citep{Weidner13b}.
The lower limit of the embedded cluster mass is assumed to be $M_{\rm ecl}^{\rm min}=5\,M_\odot $, based on the smallest observed embedded  clusters in e.g. the Taurus-Auriga star-forming region \citep{Kirk12,Kroupa03b, Joncour18}. The maximum mass of the embedded clusters, $ M_{\rm ecl}^{\rm max} $, in the forming UCD can be determined together with the normalization constant, $k_{\rm ecl}$, by solving  the following equations \citep{Weidner04b}:    
\begin{equation}
\begin{aligned}
    M_{\rm init}=\ \int_{5 \msun}^{M_{\rm ecl}^{\rm max}} M_{\rm ecl} \xi_{\rm ecl}(M_{\rm ecl}) dM_{\rm ecl}, \\
    1=\ \int_{M_{\rm ecl}^{\rm max}}^{10^{9} \msun} \xi_{\rm ecl}(M_{\rm ecl}) dM_{\rm ecl}.\\
	\label{eq:meclmax}
\end{aligned}
\end{equation}
The upper integration limit of $10^9 M_\odot$ is assumed to be the upper physical limit of  embedded cluster masses where  star cluster-type systems appear to end  \citep{Dabringhausen08}, but it is not clear if such an upper mass limit exists and taking a larger value has a negligible effect on the results.

Therewith the composite IMF of the UCD, $\xi_{\rm comp}(m)$,  and its initial mass in stars, $M_{\rm init}$, are determined (Eq.~\ref{eq:mergeIMF} and~\ref{eq:meclmax}, respectively).

\subsection{Stellar evolution}

We used the latest stellar evolution tracks from the Padova group \citep{Marigo07,Marigo08} to calculate the time evolution of the physical properties of the stellar population. Assuming that the whole stellar population forms simultaneously, the same metallicity is adopted for all stars. At the endpoints of stellar evolution, stars  become remnants such as white dwarfs (WDs), neutron stars (NSs) or black holes (BHs), depending on the initial mass. We assigned remnant masses to dead stars based on the initial masses of stars following the \citet{Renzini93} prescription: stars with initial masses $0.85\msun < m_i < 8.5\msun$ leave a WD of mass $0.077 m_i +0.48$, initial masses    $8.5\msun \leq  m_i < 40\msun$  leave behind a $1.4 \msun$  NS, and initial masses $m_i \geq 40\msun $ leave a remnant BH of mass  $0.5 m_i$. Using this prescription for the masses of remnant stars is standard in stellar population synthesis modeling \citep{Maraston98, Bruzual03}.  It should be mentioned that the remnant mass correlates with the metallicity such that a lower metallicity leads to a larger remnant mass because of inefficient mass loss through stellar evolution \citep{Belczynski18} which probably leads to a higher  $M/L_V$ ratio at lower metallicity. However, the dependence of the remnant mass on the metallicity and other physical parameters are not considered here  and needs further investigation.

The contribution of both stars and remnants is accounted for in the calculation of mass and in the $M/L_{\rm V}$ ratio of the stellar population. The stellar remnants  do not contribute  to the luminosity of the population, but their contribution can potentially be very important for the present-day dynamical mass, $M$,  of the UCD.

The fraction of remnant mass increases with the age of a simple stellar population. The rate of this increase depends on the metallicity and the shape of the IMF because the IMF becomes increasingly top-heavy with decreasing metallicity.
\section{Results}

We set up a grid of different initial UCD stellar masses and metallicities. The initial stellar mass varies in the range  $10^5 \le M_{\rm init}/M_\odot \le 10^9$. We assume three different values for metallicity, $ Z \in \{0.0004, 0.004, 0.02\}$ and also apply the mass-dependent metallicity of \citet{Zhang_2018} in Sec. 3.1.3.  Assuming different ages for UCDs  from 5 to 12.5 Gyr, the physical properties of modeled UCDs (the $M/L_{\rm V}$ ratio and the half-mass radius) that have formed according to the monolithic and merging scenarios are computed and are compared with observations in this section.

\subsection{Mass-to-light ratio}

The dynamical mass of a stellar system with a radius which is much smaller than the tidal radius, $r_t$, evolves through stellar and  dynamical evolution. However, the dynamical evolution is not a dominant process for UCDs with a two-body relaxation time longer than a Hubble time \citep{Mieske08, Baumgardt-Mieske08}.
In addition, for a stellar system with initial mass larger than, for example,  $M_c=10^6 M_\odot$, evolving on a circular orbit with a Galactocentric radius of about $R_G=100$ kpc in a Milky-Way-type galaxy, the tidal radius is larger than 500 pc which implies that almost all stars remain bound\footnote{We use $r_t=R_G\left(\frac{M_cG}{3V^2_{f}R_G}\right)^{1/3}$ assuming $V_{f}=$200 km/s for the asymptotic rotational speed of the MW.}. Even in a very dense galaxy cluster environment such as NGC 1399, the central galaxy of the Fornax cluster, only about 10\% of UCDs exhibit probable signatures of tidal tails and general asymmetric features \citep{Voggel_2016}.

The $M/L_{\rm V}$ ratio of a GC and UCD depends on the retention fraction of dark remnants and  natal kicks that NSs and BHs receive during supernova explosions (WDs are assumed to receive no kick).  \cite{Peuten16} and \cite{Baumgardt17} use the observed degree of mass segregation in GCs to infer using N-body models that the retention fraction in these is less than 50\%.    The actual retention fraction in UCDs is still a matter of debate \citep{Jerabkova17, Pavlik18}, because studying the  close dynamical encounters in systems as massive as UCDs is computationally expensive and is beyond the present-day computational capability. For massive UCDs  with birth masses larger than $10^7 M_{\odot}$, the gravitational potential is sufficiently deep such that remnants with a velocity kick dispersion up to $\sigma_{BH}=$300 km$s^{-1}$  are not likely to escape (see \citealt{Jerabkova17} for explicit calculations).  This implies  that the upper limit for the  retention fraction can be $100\%$ for very massive UCDs \citep{Jerabkova17, Pavlik18}. To cover the whole possible range we assume that all stars form at the same time, and  we calculate in Sec.~\ref{RF10-100} the stellar plus remnant mass, $M$, and the V-band luminosity, $L_{\rm V}$,  for two different values of the retention fraction of $10\%$ and $100\%$ \citep{Pavlik18}. In Sec.~\ref{sec:fret} the more realistic case that the retention fraction depends on the initial UCD mass is included following the formulation by \cite{Jerabkova17}. In Sec.~\ref{SpatialBH} the timescale is estimated for the BHs to dynamically segregate to a compact central sub-cluster.


\subsubsection{Fixed retention fraction of 10\% and 100\%}\label{RF10-100}

The V-band luminosities of the UCD models are plotted in Fig.~\ref{fig:LV} for different formation scenarios as well as for an invariant canonical IMF.
As can be seen, the monolithic collapse models are less luminous  for massive UCDs compared   to the merging models.  This is because in the monolithic collapse scenario,  the more massive UCDs have a more top-heavy IMF leading to a larger population of $\rm O$ and $\rm B$ stars that quickly turn to dark remnants.

\begin{figure*}
\begin{center}
\includegraphics[width=125mm,height=103mm]{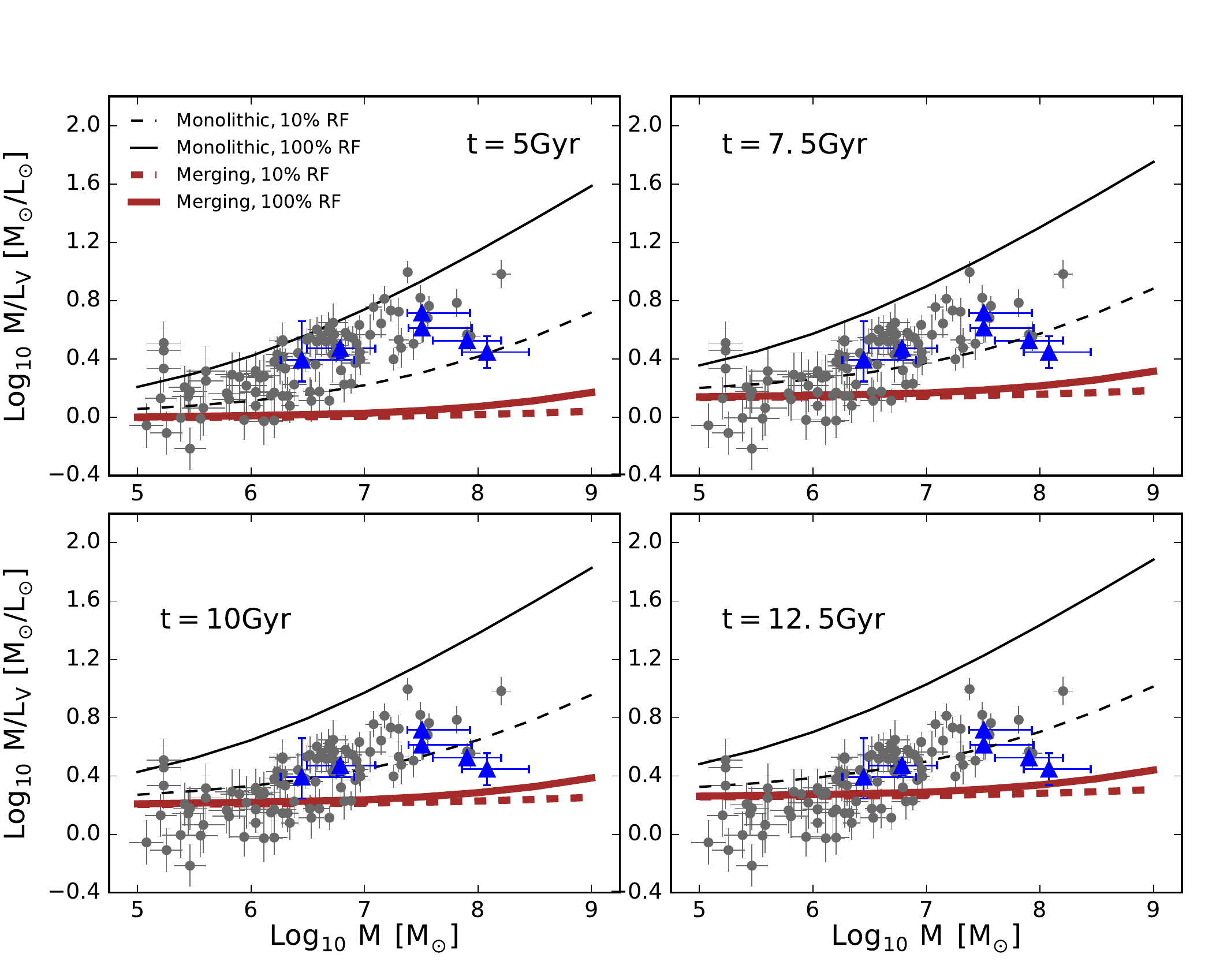}
\caption{The $M/L_{\rm V}$ ratio vs. present-day dynamical mass of modeled UCDs at different ages is compared with observational data plotted as gray circles \citep{Mieske08, Mieske13}. The UCDs that are thought to have SMBHs are shown with blue triangles (see Sec.  \ref{SpatialBH} for more details). Assuming a 10\% (dashed lines) and 100\% (solid lines) retention fraction (RF) of dark remnants, the $M/L_{\rm V}$ ratio is  calculated for both monolithic and merging scenarios shown as black and red lines, respectively. For all models we assume $Z=0.0004$.
}
\label{fig:realistics}
\end{center}
\end{figure*}

Fig.~\ref{fig:realistics} presents  the  $M/L_{\rm V}$ ratios of modeled UCDs versus their mass based on the monolithic  (black lines) and merging scenario (red lines).
In the monolithic collapse models, the $M/L_{\rm V}$ ratio increases with the mass of the UCDs. This trend is more pronounced for models with a 100\% retention fraction. In the merging scenario, the $M/L_{\rm V}$ ratio remains almost constant with the stellar mass of the UCDs. The reason for this difference is that they follow different IMFs as described in Section 2.
Also, in the merging scenario, the retention fraction doesn't play an important role on the present-day $M/L_{\rm V}$ ratio of UCDs.  This is because in the merging scenario, the retention fraction is computed for each individual cluster such that the UCD-wide retention fraction will be much smaller than in the monolithic case when the mass-dependent retention fraction is used.Therefore, the number of remnants in the merging scenario is less than what is expected in the monolithic collapse models in massive UCDs.

As can be seen in Fig. \ref{fig:realistics},  most of the observed data lie between the calculated $M/L_{\rm V}$ ratios of the merging and monolithic models. However, the observed data show a trend with the stellar mass of the UCDs which is in better agreement with the results of the monolithic collapse scenario.   We conclude that the observed UCDs with large $M/L_V$ values may be monolithically formed, while those with small values may be merged star-cluster complexes. Those in between would be a mixture of a monolithically formed component which merged with other star clusters in a centrally concentrated cluster complex as are observed in, for example, the Tadpole galaxy's tidal tail (fig. 10 in \citealt{Kroupa15}).

\begin{figure*}
\begin{center}
\includegraphics[width=125mm,height=103mm]{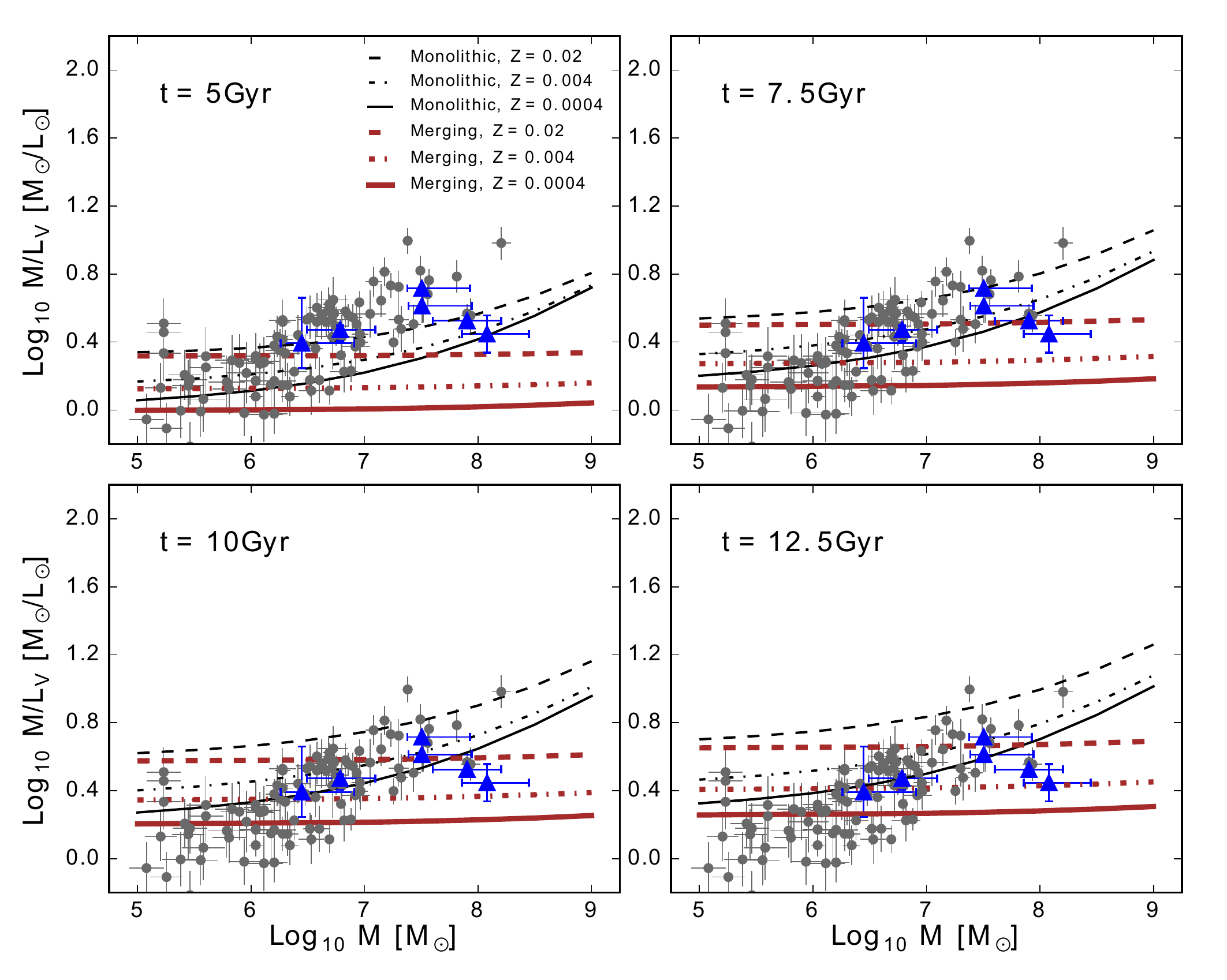}
\caption{As Fig. \ref{fig:realistics}, but for different values of the metallicity.  The retention fraction of dark remnants is assumed to be 10\% for all models.
}
\label{fig:M/Lv4}
\end{center}
\end{figure*}

Fig.~\ref{fig:M/Lv4} shows the effect of the metallicity on the $M/L_{\rm V}$ ratio of the UCDs as based on the monolithic collapse and merging scenarios.  The $M/L_{\rm V}$ ratios decrease with decreasing $Z$ because a lower-metallicity stellar population is brighter than a metal-rich one of the same mass.

\subsubsection{Mass-dependent retention fraction}
\label{sec:fret}

Here, we consider the dependency of the retention fraction of NSs and BHs on the birth masses of UCDs. The potential of massive UCDs is  too deep to allow remnants to escape while almost all dark remnants escape from low mass systems. This dependency is studied by \citet{Jerabkova17} assuming different kick velocities for dark remnants. The following functional form of the retention fraction-mass scaling relation is the best-fit to the data in fig.~2 of \citet{Jerabkova17} for  $\sigma_{\rm kick}=300\,$km/s for NSs and BHs:
\begin{equation}
RF= 1- \frac{1}{1+e^{a\log_{10}(M_{\rm init}/M_0)}},    
\label{equ:fit}
\end{equation}
where, the fitting parameters are $a=3.7$, and $M_0=1.95\times10^7 M_{\odot}$. According to \citet{Jerabkova17}, adopting $\sigma_{\rm kick}=300\,$km/s is reasonable to assume that the retention fraction of UCDs with birth masses larger than $10^7 M_{\odot}$ is close to 100\%. A smaller values of $\sigma_{\rm kick}$ gives a higher retention fraction and consequently leads to a higher $M/L_V$ ratio.

The retention fraction is estimated by assuming that 10\% of all NSs and BHs are retained (Eq.  \ref{equ:fit}) in a GC with mass $10^5 M_{\odot}$ (cf. \citealt{Baumgardt17}). With this normalization condition and using the $r_{\rm h_0}(M_{\rm ecl})$ relation (Eq.~5), the retention fraction is calculated for stellar remnants as a function of a UCD's initial  mass ($M_{\rm init}$).  This relation was used throughout the modelling because most of the BHs are produced within about 5-10 Myr and during this time the young UCD does not expand significantly as the mass loss is not significant yet.

A mass-dependent retention fraction can play a significant role in the monolithic collapse scenario owing to the  top-heaviness of the IMF. However, in the merging scenario the IMFs of individual clusters that merge to form a UCD are not as top-heavy as those in monolithically formed UCDs. In the merging scenario, the retention fraction is computed for each individual cluster such that the UCD-wide retention fraction will be much smaller than in the monolithic case when the mass-dependent retention fraction is used.  Therefore, different retention fractions have negligibly different consequences on the $M/L_{\rm V}$ ratio in the merging scenario.


\begin{figure*}
\begin{center}
\includegraphics[width=55mm,height=120mm]{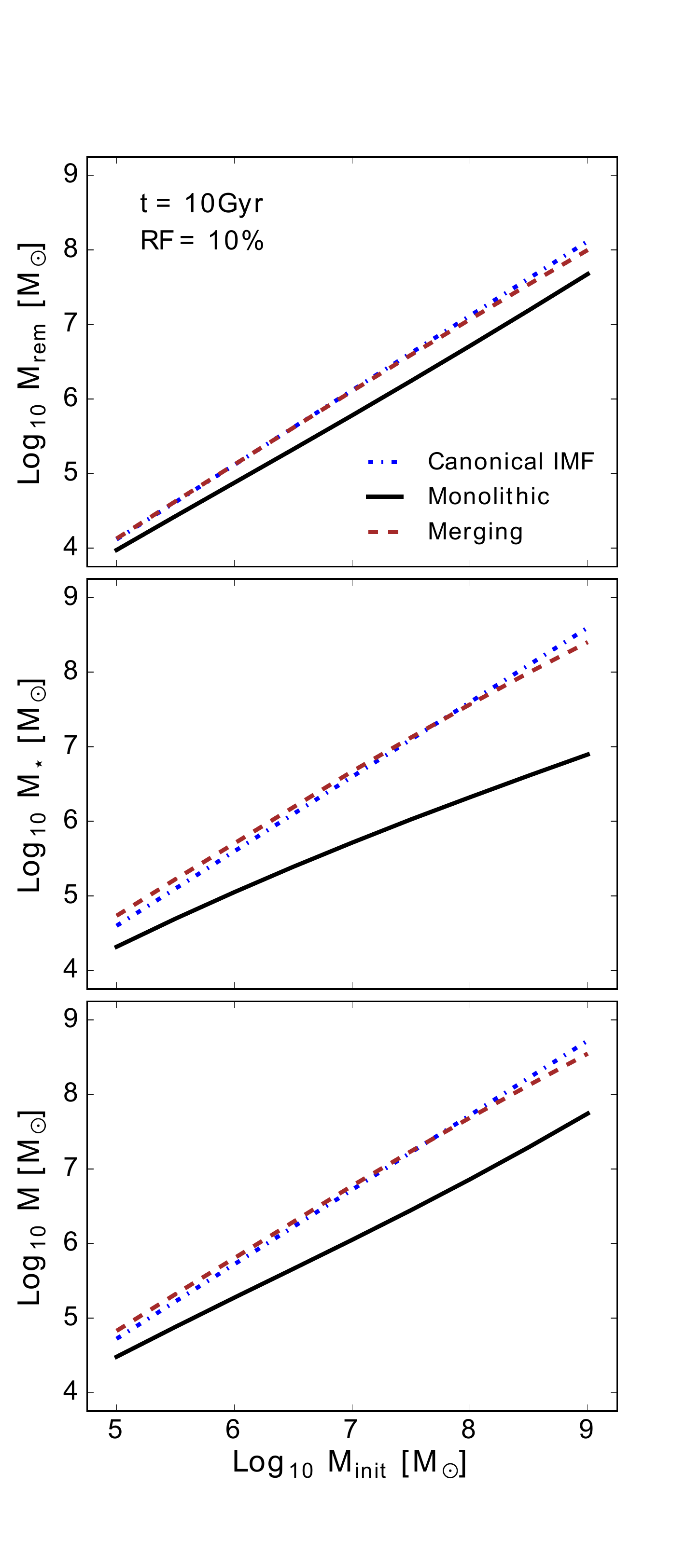}
\includegraphics[width=55mm,height=120mm]{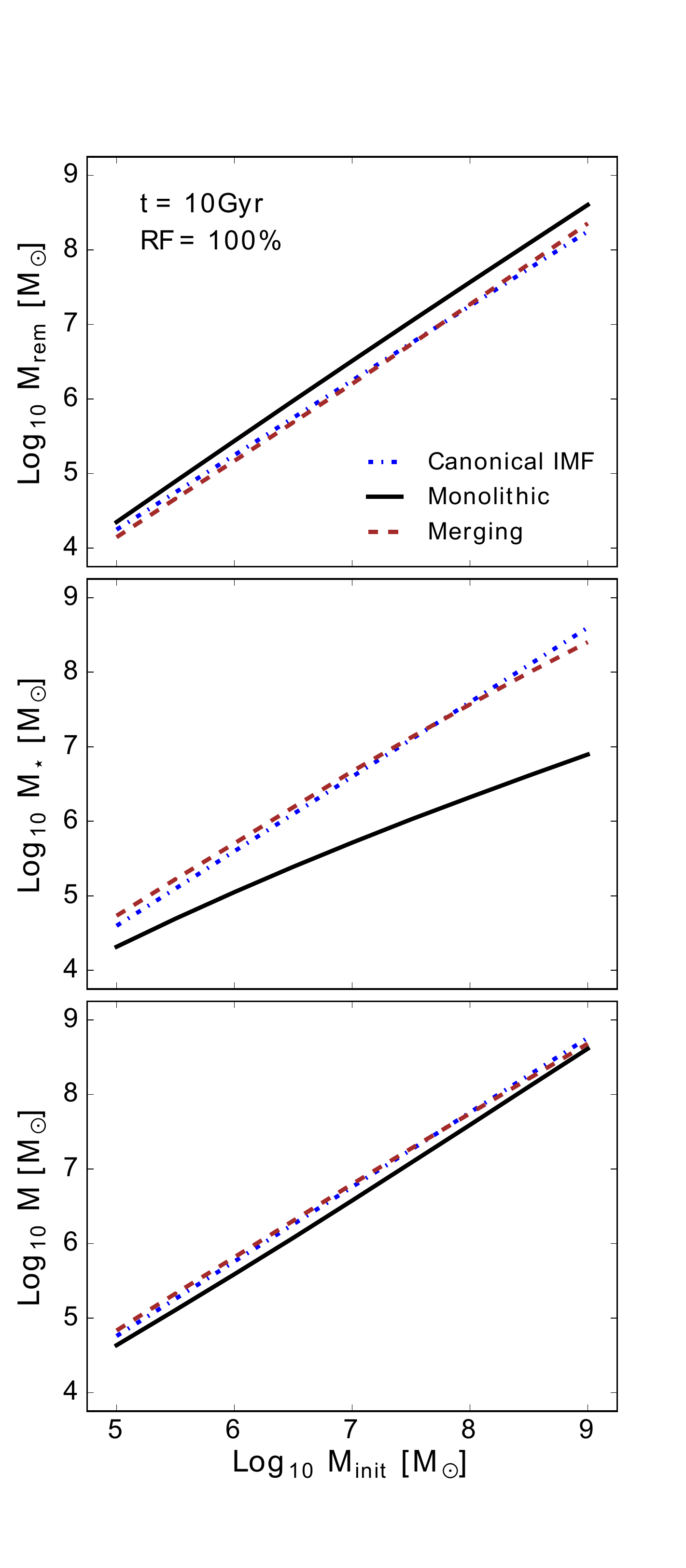}
\includegraphics[width=55mm,height=120mm]{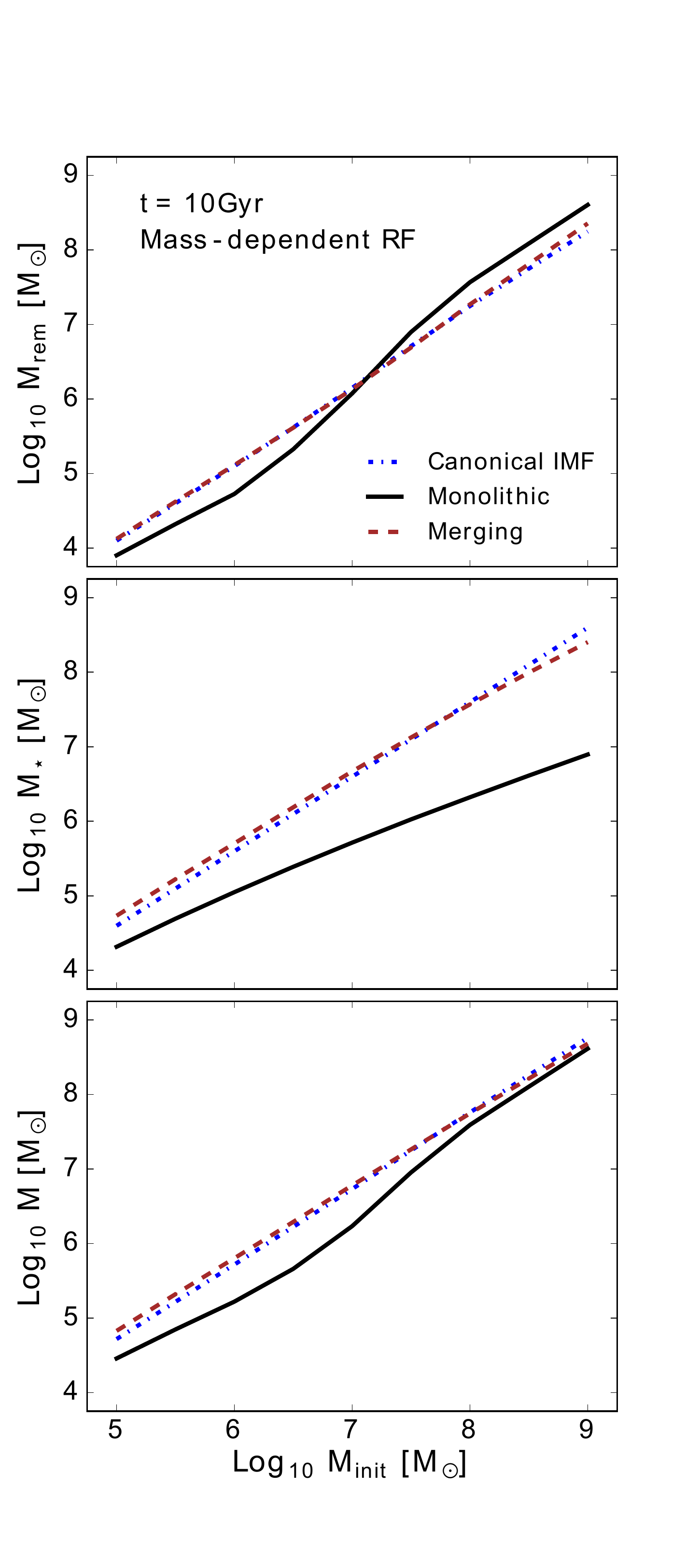}
\caption{The SPS-calculated stellar mass, $M_*$, remnant mass, $M_{\rm rem}$, and total mass (live stars $+$ remnants), $M$, of UCDs after 10 Gyr of evolution in different formation scenarios is plotted as a function of the initial stellar mass, $M_{\rm init}$. Three different retention fractions (RF) for remnants are adopted. The dependency of RF on the initial  mass of the UCDs (Eq. \ref{equ:fit}) is adopted from \citet{Jerabkova17} for a kick velocity distribution of  $\sigma_{\rm kick}=$300 km/s.
}
\label{fig:fraction1}
\end{center}
\end{figure*}

 In Fig.~\ref{fig:fraction1} we compare the mass of remnants,  live stars and the total  mass (live stars plus remnants) of UCDs for different formation scenarios as well as for an invariant canonical IMF adopting  $RF=$10\%, 100\% and the mass-dependent retention fraction (Eq. \ref{equ:fit}).

Adopting a 10\% retention fraction, the mass of remnants is lower for the case of the monolithic collapse model compared to the  merging scenario and invariant canonical IMF model. Indeed, models with an initially  top-heavy IMF create and retain more BHs and NSs, while  more WDs are created and finally retained in the cases with a canonical IMF \citep{Haghi2020}. For models with  $RF=10\,$\%, the total mass that remains in  the UCDs is lower for monolithic collapse models. This is because in the monolithic collapse model, a large fraction of mass is converted into NSs and BHs of which only 10\% remain in the UCD.  Therefore, the contribution of BHs and NSs that remain in a UCD decreases.

\begin{figure}
\begin{center}
\includegraphics[width=85mm,height=170mm]{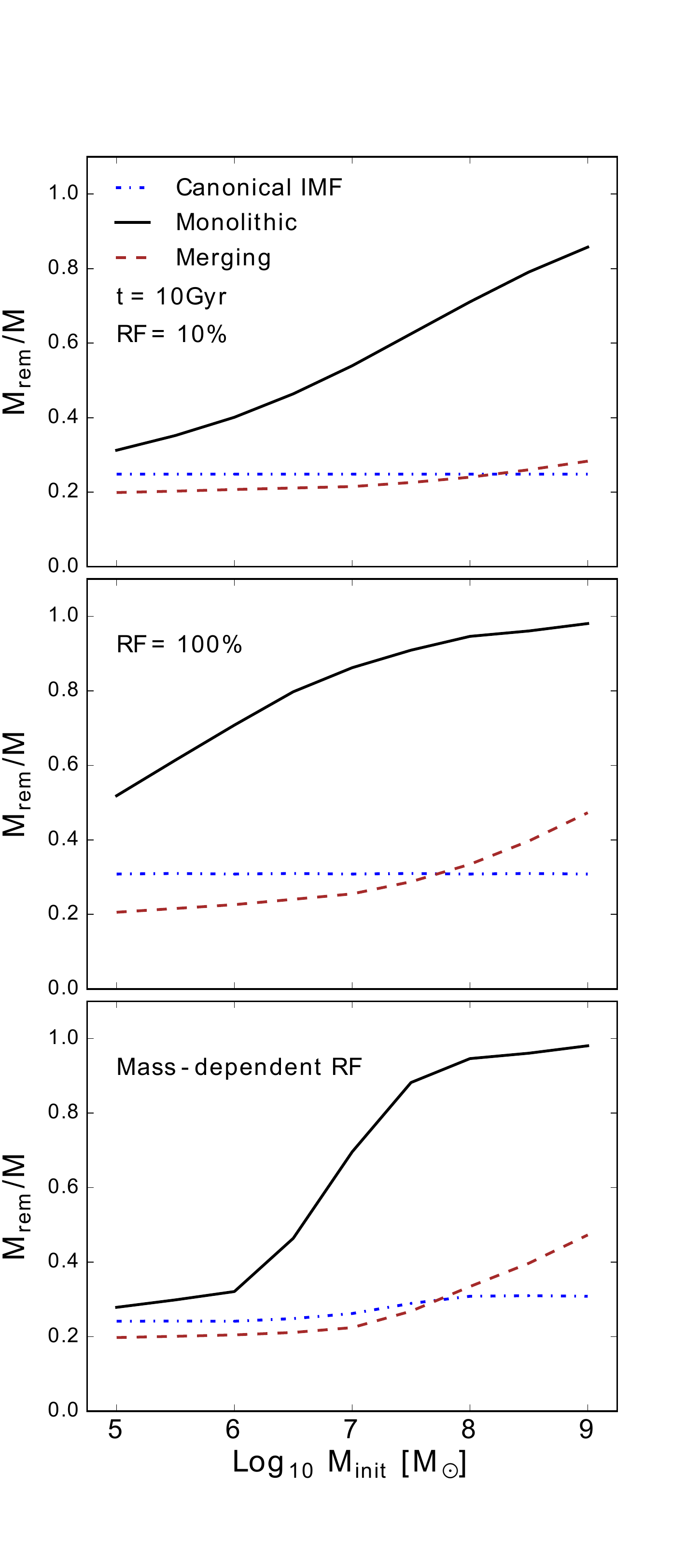}
\caption{The fraction of mass in remnants, $M_{\rm rem}/M$, at $t= $10 Gyr is plotted in dependence of the initial UCD mass, $M_{\rm init}$,
for different formation scenarios assuming the retention fraction of remnants being 10\% (top panel), 100\% (middle panel) and mass-dependent (bottom panel).
}
\label{fig:fraction2}
\end{center}
\end{figure}

In order to show which fraction of mass is in the remnants we plot, in Fig. \ref{fig:fraction2}, $M_{\rm rem}/M$ as a function of the initial mass of the UCDs for the different formation scenarios adopting the three different retention fractions. As can be seen, the fraction of mass in remnants is higher in the monolithic collapse scenario compared to the merging models.

\begin{figure*}
\begin{center}
\includegraphics[width=115mm,height=98mm]{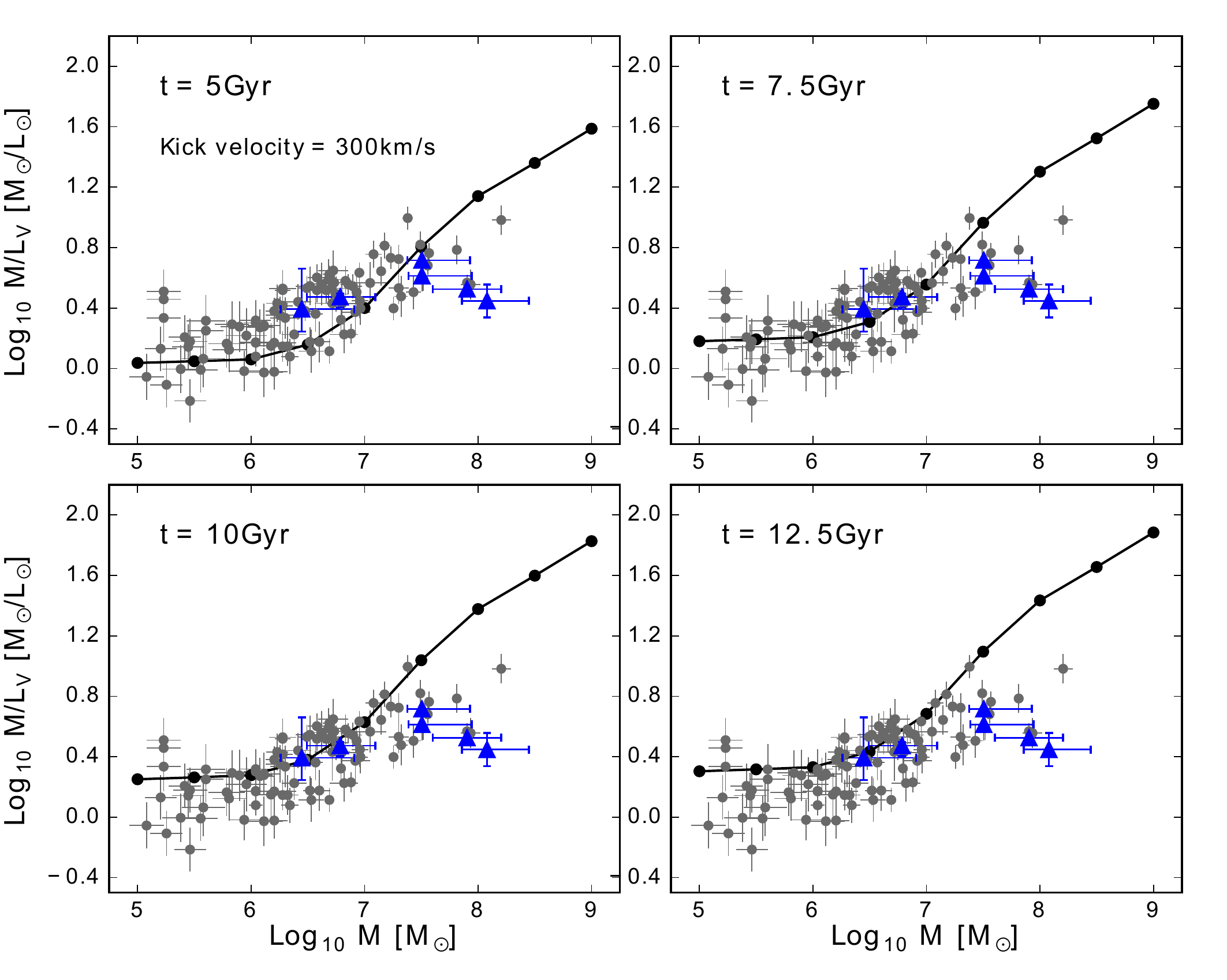}
\caption{ The SPS-calculated $M/L_{\rm V}$ ratios of UCDs in dependence of their present-day dynamical mass, $M$,  constructed based on the monolithic collapse scenario, adopting  the mass-dependent retention fraction (Eq. \ref{equ:fit}) of dark remnants (black filled circles joined by the solid black line).  Assuming four different ages for the UCDs, the resulting $M/L_{\rm V}$ ratios are  compared with observational data that are as in Fig. \ref{fig:realistics}. The observed data are in better agreement  with the model $M/L_{\rm V}$ values  at ages between 7.5 and 10 Gyr. For all models we assume $Z=0.0004$.
}
\label{fig:variRF}
\end{center}
\end{figure*}

The influence of the mass-dependent retention fraction on the $M/L_{\rm V}$ ratio is shown in Fig.~\ref{fig:variRF}. For all modelled UCDs the monolithic collapse scenario with  $Z=0.0004$ is adopted. As can be seen, adopting more realistic conditions for the  retention fraction improves the agreement between the observed $M/L_{\rm V}$ ratios  and those resulting from the monolithic scenario. Ages of UCDs in the range of 7.5 Gyr are consistent with the observational data which show UCDs are mostly old systems \citep{Chilingarian11}. In the next section we show how applying the observed mass-metallicity relation of UCDs improves the consistency of the models with observed UCDs data.

\begin{figure*}
\begin{center}
\includegraphics[width=115mm,height=98mm]{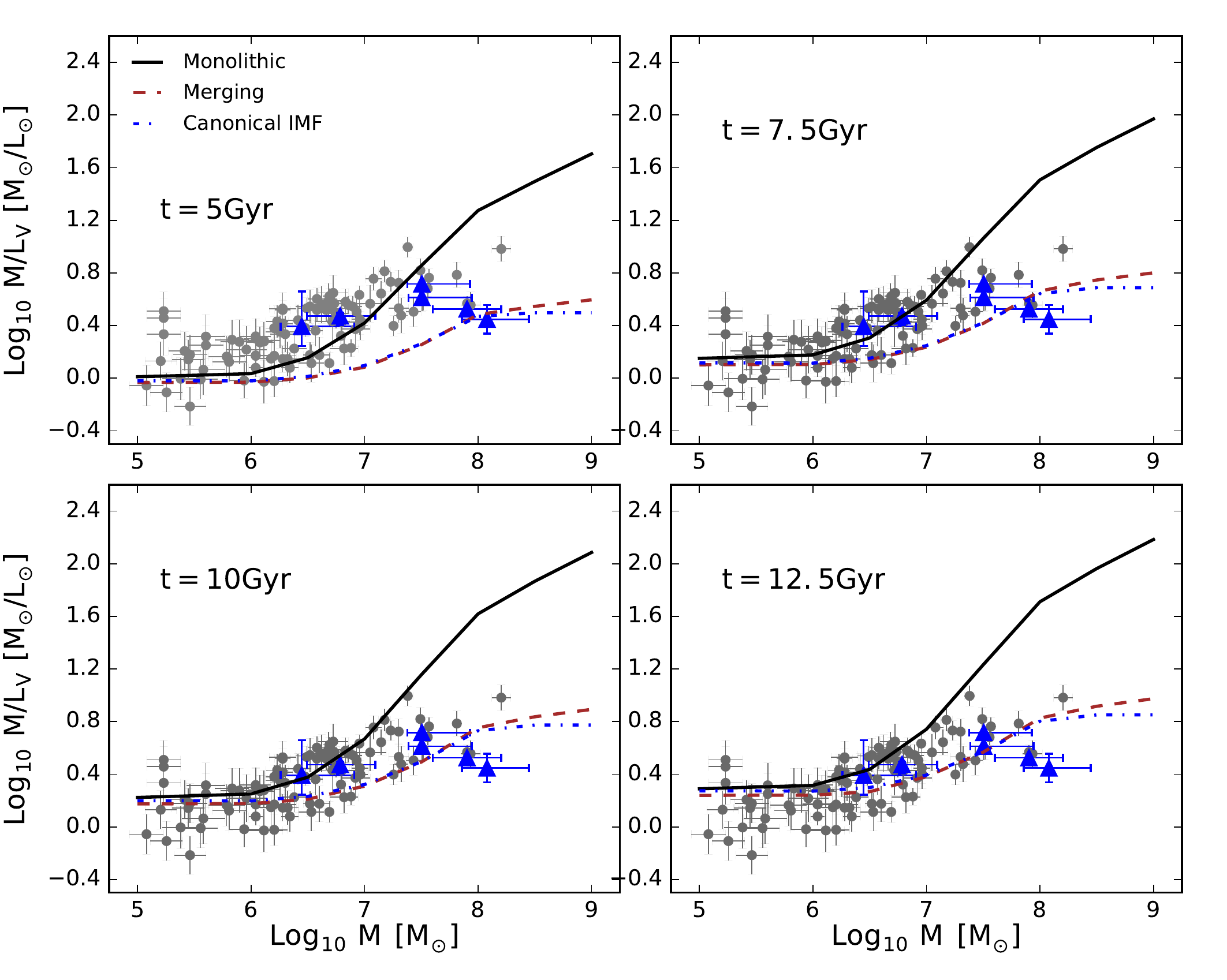}
\caption{ The SPS-calculated $M/L_{\rm V}$ ratios of UCDs as a function of their present-day dynamical mass assuming the mass-metallicity relation (Eq. \ref{MZrelation}) and mass-dependent retention fraction (Eq. \ref{equ:fit}) for different formation mechanisms. The resulting $M/L_{\rm V}$ ratios assuming the canonical IMF, monolithic collapse and the merging scenario are shown as blue, black and red lines, respectively. The data are as in Fig. \ref{fig:realistics}.
}
\label{fig:Z_RF_mass_dependent}
\end{center}
\end{figure*}

\begin{figure*}
\begin{center}
\includegraphics[width=115mm,height=98mm]{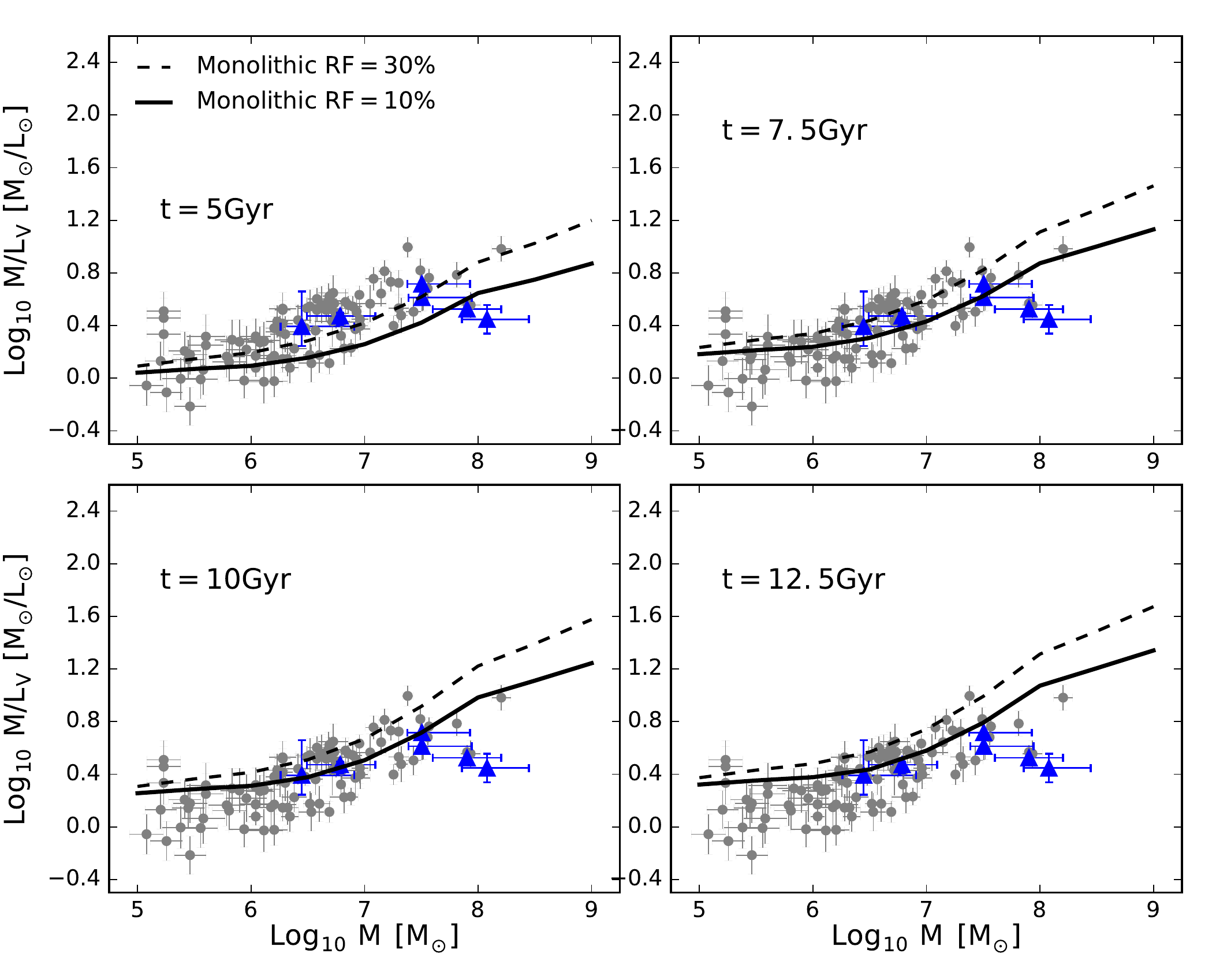}
\caption{ The SPS-calculated $M/L_{\rm V}$ ratios of UCDs as a function of their present-day dynamical mass assuming the mass-metallicity relation (Eq. \ref{MZrelation}) and a constant retention fraction for the monolithic collapse scenario. The resulting $M/L_{\rm V}$ ratios assuming RF=10\% and 30\% are shown as solid and dashed lines, respectively. The observational data are as in Fig. \ref{fig:realistics}. The observed data are in better agreement  with the model $M/L_{\rm V}$ values (with RF=10\%) at ages between 7.5 and 10 Gyr.}
\label{fig:Z_dependent_RF_cte}
\end{center}
\end{figure*}

\begin{figure}
\includegraphics[width=\columnwidth]{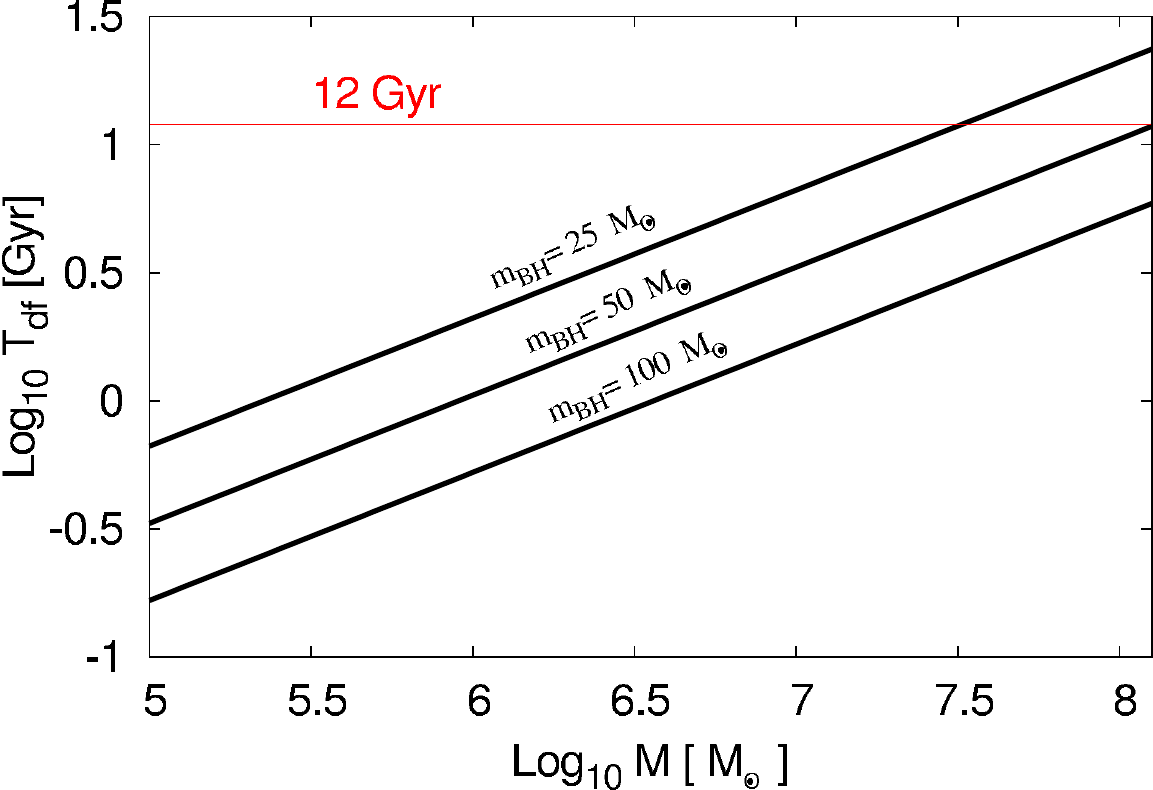}
\includegraphics[width=\columnwidth]{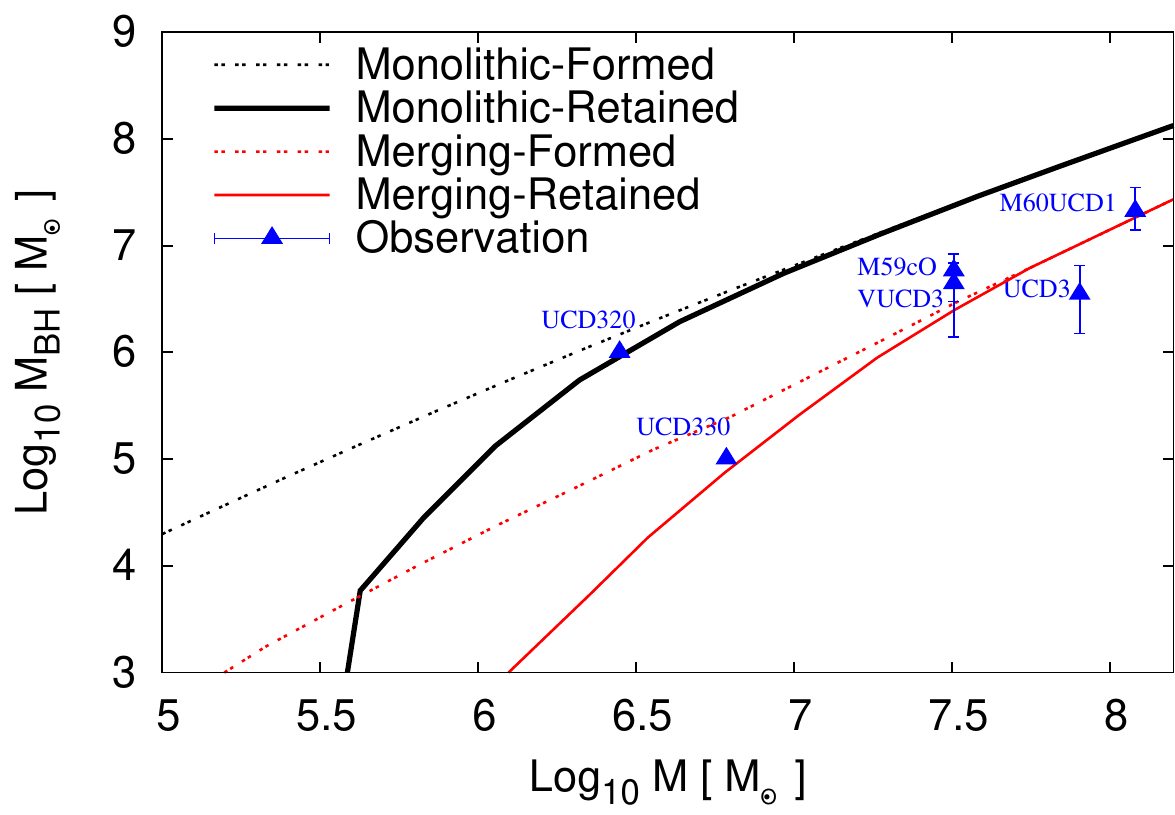}
\caption{Upper panel: The dynamical friction time-scale vs. present-day dynamical mass of the UCDs. Assuming the BHs have a mass of 25, 50 and 100 $M_{\odot}$, the time scale for the BHs to sink to the core of their UCD through dynamical friction is calculated using Eq.~\ref{eq:DF}. The red horizontal line shows $T=12$ Gyr. Lower Panel: The total mass of BHs that is formed and retained in UCDs with different masses for merging and monolithic collapse scenarios assuming the mass-dependent retention fraction (Eq. \ref{equ:fit}) and the mass-metallicity relation (Eq. \ref{MZrelation}) which enters the value of $\alpha_3$ (Eq. \ref{eq:alpha3}).  The six known central SMBH masses and the  dynamical masses of their hosting UCDs taken from \citet{Seth2014, Ahn_2017, Afanasiev_2018} and  \citet{Voggel_2018} are over-plotted as filled triangles for comparison.}
\label{fig:TDF}
\end{figure}

\subsubsection{Mass-dependent metallicity}

So far in this paper, we have adopted a constant metallicity for the modeled UCDs.   \cite{Zhang_2018} derived  stellar population parameters for a representative sample of UCDs  near the central galaxy M87 of the Virgo galaxy cluster and found the data to be representable with the  mass-metallicity correlation,
\begin{equation} \label{MZrelation}
    \log_{10}[Z/H]=-10.17+1.32\log_{10}[M_*/M_{\odot}],
\end{equation}
where, $M_*$ is the present-day stellar mass of UCD. Since the deeper potential well of more massive systems means a higher retention efficiency of self-enriched gas that may be incorporated into the nearly simultaneous formation of low-mass stars, such a correlation between mass and metallicity could be understood as a results of a mass-dependent self-enrichment (e.g. from supernova ejecta and stellar winds) within the proto-cluster clouds \citep{Zhang_2018}.

Given the dependence of observed dynamical  mass-to-light ratios on metallicity (as shown in Figure \ref{fig:M/Lv4}), and that UCDs generally follow a mass-metallicity relation (Eq. \ref{MZrelation}), we recalculated the $M/L_V$ values of UCDs in the context of the invariant canonical IMF, monolithic collapse and the merging scenario. Incorporating the mass-metallicity relation in $\alpha_3$ (Eq. \ref{eq:alpha3}) and with our SSP models enables us to establish a new relation between the
$M/L_V$ ratio and mass of a UCD.

In Fig. \ref{fig:Z_RF_mass_dependent} we compare the calculated  $M/L_V$ ratio for different formation scenarios. The introduction of the mass-metallicity relation into our models (which means adopting higher metallicities for massive UCDs) does not affect our previously-reached conclusions.  A comparison between Figs. \ref{fig:M/Lv4} and \ref{fig:Z_RF_mass_dependent} reveals that the incorporation of the mass-metallicity relation actually improves the consistency of the merging scenario with the observed trend.  Moreover, as can be seen, ages of UCDs between 5 and  12.5 Gyr (lower-right panel of Fig. \ref{fig:Z_RF_mass_dependent}) are well consistent with the data, in agreement with UCD ages of \citet{Chilingarian11}.  Thus by including the mass-dependent retention fraction and the mass-metallicity relation, the observational data with larger $M/L_V$ ratios are explainable as monolithically formed UCDs with ages older than 5 Gyr, which the data with lower values of $M/L_V$ ratios are also explainable as merged cluster complexes with ages older than 5 Gyr.

It is worth mentioning that given none of the observed UCDs have $M/L_V>10$, the very high $M/L_V$ ratios ($>10$) achieved for massive UCDs in the monolithic scenario show a steeper trend with increasing mass than the observed UCDs. Unless the observed UCDs with $M>10^8 M_{\odot}$ were excessively young, it seems a merging scenario can better explain the highest mass objects, while both scenarios could explain lower mass UCDs.  Comparing the $M/L_V$ results in Figs. \ref{fig:M/Lv4} and \ref{fig:Z_RF_mass_dependent} reveals that the monolithic scenario might require a constant retention fraction  to fit the whole mass range, but the merging scenario requires a mass-dependent RF. In order to evaluate this issue, a constant RF is adopted in combination with the mass-metallicity relation in Fig. \ref{fig:Z_dependent_RF_cte}. As can be seen, the SPS-calculated $M/L_{\rm V}$ ratios of UCDs assuming RF=10\% is in good agreement with the observed data for the monolithic collapse scenario at ages between 7.5 and 10 Gyr.

\subsubsection{Spatial distribution of the stellar remnants} \label{SpatialBH}

As a result of dynamical friction, the BHs that formed through stellar evolution of the most massive stars  segregate into the UCD's  core and form a sub-system of BHs. The formation of such a BH subsystem is already well studied in star clusters \citep{Banerjee10} and is also of interest in galactic nuclei \citep{Kroupa20}.
Moreover, it is shown that the star-burst cluster is most likely to form primordially mass segregated \citep{Haghi15, Pavlik19}. If not, it will mass-segregate and  the BH sub-cluster shrinks due to dynamical mass segregation to about 10\% of the initial radius of the cluster but on a Gyr time-scale \citep{Breen13, Kroupa20}.

Here we estimate the time taken by the BHs to inspiral to the UCD's core using the dynamical friction time-scale \citep{Celoria, Gerhard01}

\begin{equation}
 T_{df}=\frac{1.17~~ r_i^2v_c}{\rm ln\Lambda~~ G\,m_{BH}},
  \label{eq:DF}
\end{equation}
where $r_i$ is the initial distance of a BH from the center of the UCD, $m_{BH}$ is the mass of the BH, and $v_c$ is the mean value of the stellar velocity dispersion of the UCD. We use the  typical value of the Coulomb logarithm $ \rm ln \Lambda \approx 15 $ \citep{Celoria, Gerhard01} for an isothermal distribution of stars.

We assume the BHs have a mass of 25, 50 and 100 $M_{\odot}$ and calculate the dynamical friction time-scale assuming they begin at the final (expanded) half-mass radius of the UCDs of different mass. The final half-mass radius stabilises after about 100 Myr when  the young UCD has expelled its residual gas and after mass loss from its evolving stars has reached a negligible amount. The initial distance between the BH and the center of its host UCD is assumed to be $r_i\approx 10$ pc which is the typical  half-mass radius of a UCD.

Fig. \ref{fig:TDF} shows the dynamical friction time-scale  vs. the present-day mass of the UCD to see for which the time-scale is shorter than their age  (adopted to be 12 Gyr as the maximum possible age). As can be seen, the time scales are typically less than 12 Gyr and these remnants will spirals towards the center of their UCDs within to form  a central BH sub-cluster.  The lines in Fig. \ref{fig:TDF} for BH masses of 50 and 100 $M_{\odot}$ start to cross the Hubble time above $M=10^8 M_{\odot}$, and for BH masses of 25 $M_{\odot}$ they start to cross the Hubble time above $M=3 \times 10^7M_{\odot}$. This means that all BHs with  masses above 25 $M_{\odot}$ will sink to the center of UCDs with $M<3 \times 10^7M_{\odot}$, while for UCDs with $M> 10^8M_{\odot}$ only BHs with $m_{BH}>50 M_{\odot}$ have enough time to sink to the central region of UCDs. Since the number of such massive BHs are rare, one can assume $M= 10^8M_{\odot}$  as an upper limit of UCD mass that can form a BH sub system. This means that the central velocity spikes measured for massive UCDs  ($M>10^8M_{\odot}$) \citep{Seth2014} are more likely due to a central remnant clusters formed if the UCD was born primordially mass segregated.



It should be noted that going towards low mass UCDs, the retention fraction decreases and the IMF-slope, $\alpha_3$,  for massive stars increases,  and hence the number of stars that can turn into BHs decreases. The lower panel of Fig. \ref{fig:TDF} depicts the total mass of BHs that is formed and retained in UCDs with different masses for the merging and monolithic collapse scenarios. This implies that even if the retained BHs sink to the central part of the low-mass UCDs, the total mass of their BHs sub-system is not enough to lead to central velocity dispersion spikes. The mean value of the total mass for those UCDs for which a central spike has been detected is about $ 7\times 10^7 M_{\odot}$ \citep{Voggel19}.

In the lower panel of Fig.  \ref{fig:TDF} we added the existing SMBH mass measurements in observed UCDs. As can be seen, the actually  observed data follow the theoretical curve,  so that  there seems to be a trend of lower mass SMBHs in lower mass UCDs as expected from the present theory.



If the monolithically-formed UCDs form mass segregated, then the central sub cluster of BHs is there from the start.  More work will be needed to quantify the stellar-dynamical properties of the BH sub-cluster. A UCD with a central sub-cluster of BHs will most likely persist in a state of balanced evolution unless gas accretion onto it squeezes it into a relativistic state where about 5\% of the BH sub-cluster collapses to a SMBH through the radiation of gravitational waves \citep{Kroupa20}.
For UCDs that formed from merged cluster complexes, these can only get a central BH sub-cluster through the above dynamical friction.


From Fig. \ref{fig:TDF} it can be seen that the merger model predicts the masses in the BHs in a UCD to be very similar to the observationally deduced SMBH masses. On the other hand, in Fig. \ref{fig:Z_RF_mass_dependent} it is evident that the observationally deduced dynamical $M/L_V$ values of these same SMBH-bearing UCDs can be explained if these present-day UCDs with a mass in the range $10^6-10^7 M_{\odot}$ formed monolithically, while the more massive UCDs with such SMBHs need to have formed as star cluster complexes (SCCs). How can this be understood?

The following reasons may be responsible:

\begin{itemize}

\item (a) An SMBH can only form, according to \citet{Kroupa20}, if gas falls onto a pre-existing cluster of BHs in order to force it out of balanced evolution and to force it to reach a relativistic state such that the BH cluster collapses through the radiation of gravitational waves to a SMBH (which, in the forming galaxy, will continue to grow through Eddington-limited accretion). Such an environment is only available at the centre of a forming major galaxy and not in a UCD. \citet{Kroupa20} suggest that UCDs which truly carry SMBHs may be ejected nuclei from the violent spheroidal galaxy formation process. In this case, the SMBH within such a UCD would be composed of at least about 5 per cent of the BH cluster. This would therefore explain why the observed SMBH-bearing UCDs have observed SMBH masses which are not the entire dark mass in the UCDs that accounts for the high observed $M/L_V$ values (the dark mass being in the form of those BHs that have not merged with the SMBH according to \citealt{Kroupa20}). This may perhaps be the origin of the SMBH-bearing UCDs with  $10^6<M/M_\odot <10^7$, but the lack of evidence for mergers playing a significant role in galaxy formation and evolution (e.g. \citet{Kroupa15, Kroupa20} and references therein) makes this less likely.

\item (b) On the other hand, it is possible that a SMBH-bearing UCD formed monolithically as a hypermassive star-burst cluster with a top-heavy IMF as described here and that it's BH content remains in the state of balanced evolution with the rest of the UCD whereby the observationally-deduced SMBH is merely the central concentrated component of the BH sub-cluster which is in the process of core-collapse. In this case the SMBH is not a true super-massive black hole but a compact sub-cluster of BHs. The core will rebound though through binary-BH formation (cf with \citealt{Giersz2003}) such that it is possible that the observable signature for the central velocity cusp may disappear, while it appears in other UCDs. This means that a UCD formed monolithically and not as the central cluster of a forming major galaxy may repeatedly show such cusps with the separation in time between these cusps being about 20 two-body relaxation times (e.g. \citealt{Giersz2003}). This may perhaps be the origin of the SMBH-bearing UCDs with  $10^6<M/M_\odot <10^7$ in which case the SMBH is, however, merely the central sub-cluster of BHs which is in the process of undergoing core collapse after which the core will rebound as described in the preceding text. The time-scale for core collapse would be too long in more massive UCDs.

\item  (c) The observed UCDs with SMBHs and $M>10^7\,M_\odot$ are more consistent with being born as merged SCCs: The observationally deduced SMBH masses are comparable to the calculated BH content of these and the dynamical $M/L_V$ values are also comparable. This would imply that these UCDs are likely to be explained as follows: The BH content is less massive in these than in the monolithically-formed UCDs of the same present-day mass. The BHs in the merged UCDs will mass-segregate and are more likely to form a more compact sub-cluster at the centre of the present-day UCDs which is observationally interpreted to be a SMBH. This opens the following problem that would be interesting to calculate: Given a luminous mass of a UCD, which radius does the BH sub-cluster within it evolve to for different masses of the BH sub-cluster? It is expected that a more massive BH sub-cluster will stabilise with a larger radius, because its shrinkage through mass segregation of the BHs on the stellar background will cease once the inner region begins to be dominated by BHs.

\end{itemize}


\subsection{Half-mass radius}

Above it has been shown that if most UCDs in the sample of \cite{Mieske08, Mieske13} were to form as merged cluster complexes, then those with high observed $M/L_{\rm V}$ ratios can neither be reproduced  with the canonical IMF nor with the IGIMF, even if the IMF of each star cluster is the metallicity and density dependent \cite{Marks12b} IMF (Eq. \ref{eq:alpha3}).  As an independent constraint, these UCD models can also be tested for agreement with the observed present-day radii of the \cite{Mieske08, Mieske13} UCDs.  The simulations by \cite{Bruns11} have indeed already demonstrated that merged cluster complexes can account for the observed radii of UCDs, provided the radius of the cluster complex is adjusted to do so. The observed properties of very young cluster complexes (e.g. as are evident in the tidal tail of the Tadpole galaxy, fig. 10 in \citealt{Kroupa15}, or in the Antenna galaxies) are broadly consistent with this.

On the other hand, for those UCDs that form monolithically with the \cite{Marks12b} IMF and the \cite{Marks12} half-mass-radius--mass relation, the observed $M/L_{\rm V}$ data can be well reproduced if also the mass-dependent  remnant retention fraction  (Eq. \ref{equ:fit}) and the mass-metallicity relation are adopted (Fig. \ref{fig:variRF}). Such UCDs would be born with very small half mass radii and it is a priori unclear if the UCDs would be able to expand sufficiently to evolve to have radii which agree with those observed. The simulations of the evolution of initially compact UCDs which undergo gas expulsion and stellar evolutionary mass loss by \cite{Dabringhausen10} already indicate that they do expand to the observed radii for a top-heavy IMF. We return to this test here, as the birth radius--mass relation  derived by \cite{Marks12}  is used in the present modelling. This means that it is assumed that UCDs form following the same relations (IMF, radius--mass relation) as the embedded star clusters in the Milky Way.

In this section we thus compare the present-day radii of UCDs in the monolithic scenario  with the observed values. For the initial half-mass radius, $r_{\rm h0}$, we use Eq.~\ref{eq:rMrel}. The initial mass-radius relation derived theoretically by \cite{Murray09} will be adopted instead in a separate study.

Four different mass-loss processes determine the present-day radii of UCDs:
\begin{enumerate}
\item Expulsion of residual gas. Above, $\epsilon=1$ has been used but smaller values are possible.  It is unclear how rapid this removal is but most likely it will be slow compared to the orbital timescales of stars in the young and still embedded UCD. In this case expansion will be a factor of $1/\epsilon$ (e.g. \citealt{Kroupa08}).
The expansion of the UCDs through residual gas expulsion is thus negligible or at most a factor of three for the canonical $\epsilon =1/3$ \citep{Lada03, Megeath_2016}.

\item  Mass-loss from stellar evolution, which initially increases the cluster radii on a short time-scale of about 100 Myr (e.g. \citealt{Baumgardt03}).  During this early phase the  evolution of massive stars is the dominant process.

\item  Mass-loss driven by two-body relaxation by which the half-mass radius of a UCD can increase  on a two-body relaxation time-scale \citep{Gieles11}. For UCDs with $M_{\rm init}>10^6\,M_\odot$ this is negligible given that the two-body relaxation time is comparable to a Hubble time \citep{Mieske08, Baumgardt-Mieske08}.

\item  Mass-loss due to a tidal field which is mainly coming from changes in the gravitational potential of the host galaxy with time and induces mass-loss through tidal shocks. Tidal stripping  tends to decrease the size of a UCD \citep{Gnedin99}.

\end{enumerate}

In order to calculate the present-day half-mass radii of the UCD models, we first consider the early expansion due to the evolution of massive stars (within 100 Myr) using the results of N-body simulations by \cite{Haghi2020} of a grid of star clusters with top-heavy IMFs. Their grid includes clusters with  different values for their initial masses and $ \alpha_{3} $ values. N-body models of clusters comprising $N>10^4$ stars are a reasonable approximation to the evolution of UCDs because the two-body relaxation time of such clusters is much longer than the time-span computed.

\begin{figure}

\includegraphics[width=91mm,height=70mm]{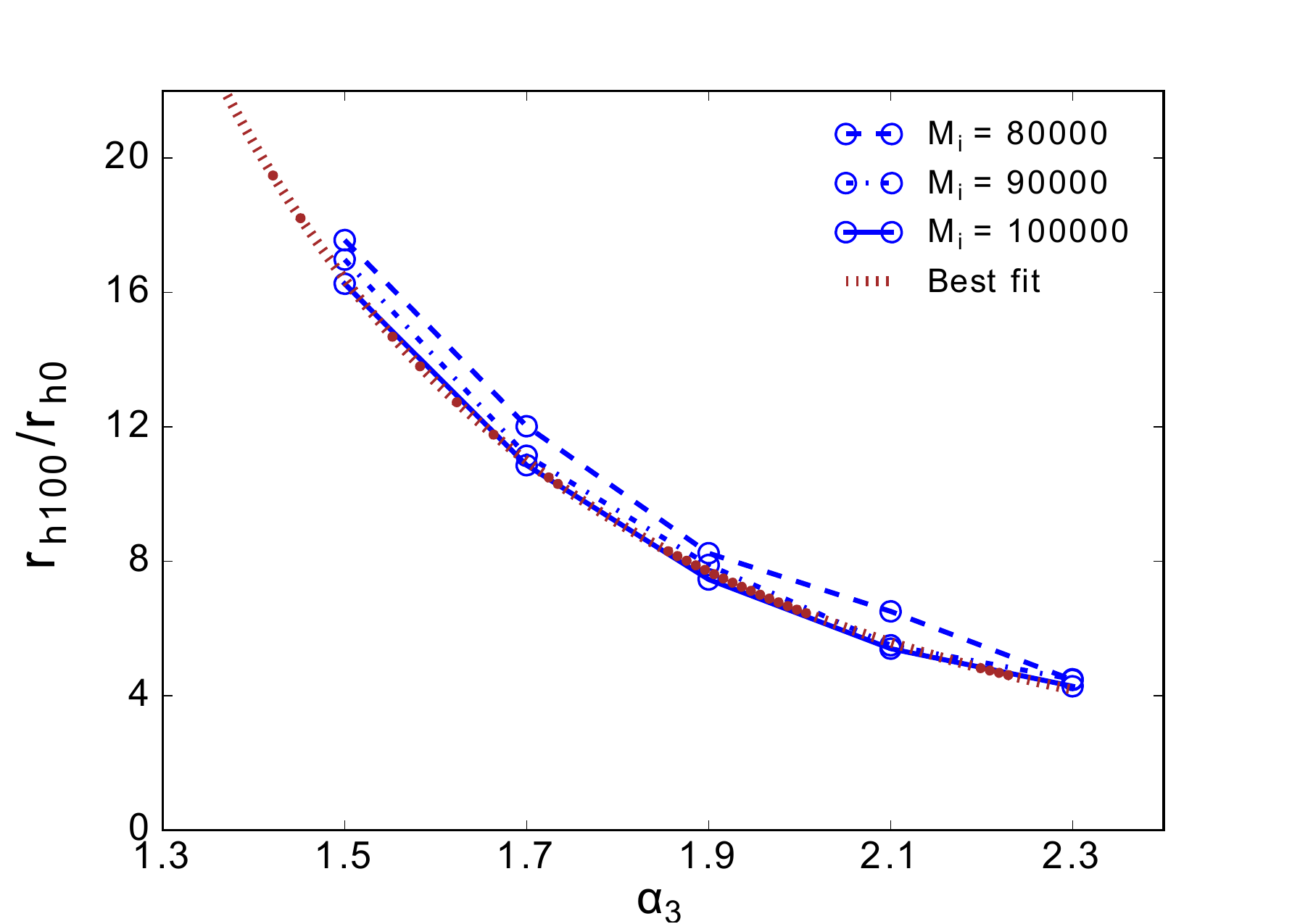}
\caption{The ratio of half-mass radius at 100 Myr  to the initial half-mass radius ($r_{\rm h100}/r_{\rm h0}$) as a function of $ \alpha_{3} $ \citep{Haghi2020}.  The blue circles indicate the results of N-body simulations of star clusters with different values of $\alpha_{3} $. The red line shows the best-fitted function (Eq. \ref{eq:bestfit}) to the radius of clusters with an initial mass of $10^5 \msun$ and different values of $\alpha_{3}$ The clusters expand more for small values of $\alpha_{3}$ due to the larger loss of mass through stellar evolution.}
\label{fig:EXP}
\end{figure}

In Fig.~\ref{fig:EXP}, we use the results by \cite{Haghi2020} and  plot the half-mass radii of the simulated star clusters at 100 Myr, $r_{\rm h100}$, versus $\alpha_{3}$ for different initial cluster masses. As can be seen, the initial mass does not play a very important role. However, for the lower values of $ \alpha_{3} $ (i.e., the more top heavy IMFs), massive stars  contribute  a large fraction to the initial mass. This leads to  more expansion due to the  evolution of massive stars.   We find the best fitting function to these data to be given by

\begin{equation}
\begin{aligned}
    \dfrac{r_{\rm h100}}{r_{\rm h0}}=60 \alpha_{3}^{-3.2}.
	\label{eq:bestfit}
\end{aligned}
\end{equation}

For each UCD model, we calculate the value of $ \alpha_{3}$ from the initial mass and adopted metallicity, following  Eq.~\ref{eq:alpha3}
(see Table~\ref{tab:alpha3} for the calculated values of  $ \alpha_{3}$).  Extrapolating Eq.~\ref{eq:bestfit}
and using the values of $ \alpha_{3} $ from Table~\ref{tab:alpha3}, the half-mass radius at 100 Myr is calculated  for different initial masses of the UCDs.

\begin{figure}
\includegraphics[width=\columnwidth]{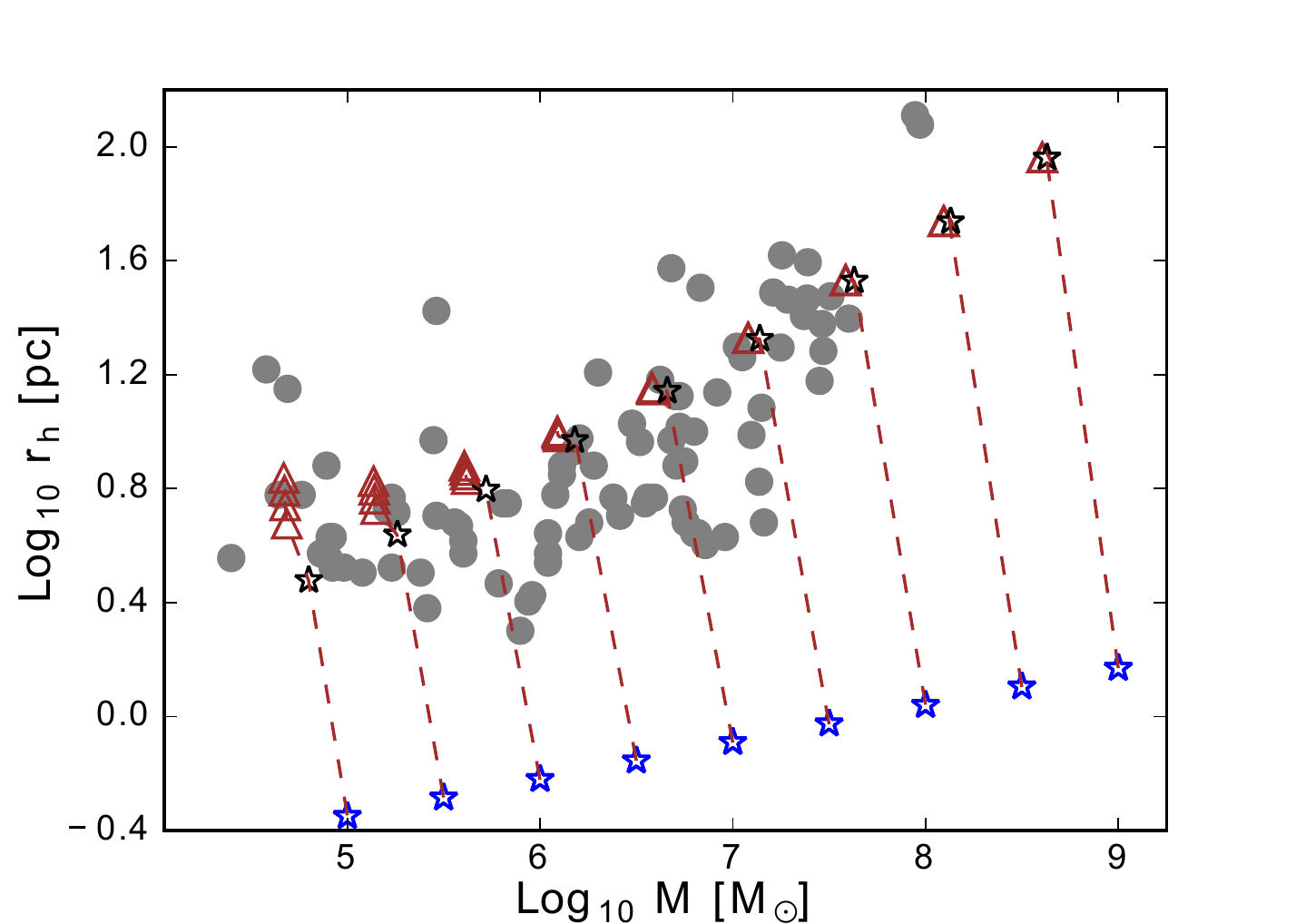}
\caption{Evolution of the radius of modeled UCDs in the monolithic collapse scenario assuming the top-heavy IMF given by Eq. 3 and  $ Z=0.0004 $. The blue stars are the initial half-mass radii, $r_{\rm h0}$, of the UCDs calculated from the \citet{Marks12} relation (Eq.~\ref{eq:rMrel}) plotted as a function of their present-day dynamical mass (See Fig. \ref{fig:fraction1}). The black stars represent their radii at 100 Myr, $r_{h100}$ (Eq. \ref{eq:bestfit}). The red triangles represent the radii, $r_{\rm ht}$, after correcting for the long-term dynamical evolution for the time $t=10\,$Gyr (Eq.~\ref{eq:rt_iso}). The dynamical mass of the UCD, $M$, is calculated for this same time $t$. The grey data points are compiled by \citet{Mieske08, Mieske13}.
}
\label{fig:RM}
\end{figure}

Finally, to calculate the present-day radius of the UCD models, the effect of long-term dynamical evolution should be considered, especially for the low-mass UCDs.
\citet{Hau09} showed that while most of the observed UCDs are in galaxy clusters, they can be assumed to be in an isolated environment because their present-day radii $<< r_{\rm t} $, where $r_{\rm t}$ is the tidal radius.  Using  the formula for the evolution of the half-mass radius of an isolated system according to \citet{Henon65}, we estimate the present-day half-mass radii of the UCD models to be given by

\begin{equation}
\begin{aligned}
    r_{\rm ht}(t)=r_{\rm h0}\ \left(1+\dfrac{t}{t_{\rm r}(0)}\right)^{2/3},
	\label{eq:rt_iso}
\end{aligned}
\end{equation}
where $t_{\rm r}(0)$ is the median two-body relaxation time of the system at $t=0$. The evolutionary tracks of the UCDs in the two-parameter plane of stellar mass and half-mass radius are plotted in Fig. \ref{fig:RM}. For all models, the monolithic collapse scenario and metallicity of $Z=0.0004$ is adopted. The model UCDs (red triangles) are in overall good agreement with the observed values.  It should be noted that the here estimated half-light radii follow the same trend as the observed values and  provide an upper envelope to the observed  present-day radii. This is because the expansion factor depends on the BH retention fraction. The  expansion factors that we used here are based on N-body models with a few percent retention fraction for the BHs. A cluster with a 100\% BH retention fraction will reach a smaller half-mass radius  (by a factor of 0.5 after about 100 Myr) compared to a model with zero retention fraction \citep{Haghi2020}. Therefore,  the lower than here computed values of the observed  half-light radii of UCDs   might be due, in the present model, to the non-zero retention fraction for the BHs and NSs. Further N-body modelling starting with a top-heavy IMF is needed to further explore this issue.

\section{Conclusions}

The possible formation mechanism of the observed UCDs is studied using the two scenarios of monolithic collapse (i.e. UCDs are the massive end of the distribution of GCs) and merging star cluster complexes. Stellar population synthesis models with a metallicity- and density-dependent IMF are constructed for this purpose.  Models based on the invariant canonical IMF are also calculated. Models with constant metallicity and with the observed mass-matallicity relation of UCDs and with constant and a remnant retention fraction which depends on the UCD mass are studied.

By varying the initial mass, metallicity and retention fraction of dark remnants,  we calculated the  dynamical $M/L_{\rm V}$ ratios and luminosities in the V-band. In this contribution we quantified how  a systematic variation of the stellar IMF can be tested with observations of  present-day UCDs.  The main conclusions can be summarized as follows:

\begin{enumerate}

\item We showed  that the SPS models with  the independently determined systematically varying IMF of \citet{Marks12b}, retaining a mass-dependent fraction of the remnants within the UCDs following \cite{Jerabkova17} and taking standard dynamical evolution into account, can explain the observed trend of $M/L_{\rm V}$ with dynamical present-day mass, $M$,  in the monolithic collapse scenario. We found that the observed $M/L_{\rm V}$  ratios are in better agreement if the ages of UCDs are about  7.5 Gyr.  Incorporating  the mass-metallicity relation of \citet{Zhang_2018} improves the agreement also to the merging scenario and leads to agreement with the observed $M/L_{\rm V}$  ratios if the ages of merged UCDs and of the  monolithically-formed UCDs are between 5 and 12 Gyr (Fig. \ref{fig:Z_RF_mass_dependent}).


\item  We found that the stellar remnants that are formed in UCDs sink to the core through dynamical friction within their ages to form  a central BH sub-cluster.  These centrally concentrated remnants are relevant for the finding of the central peak in the stellar dispersion velocities in UCDs. For the  massive UCDs ($M> 10^8M_{\odot}$) the dynamical friction timescale is larger than a Hubble time. Such  heavy UCDs can show a central peak of the velocity dispersion if the UCD formed mass-segregated such that its centre is dominated by remnants.

\item We showed that for the massive UCDs, the calculated masses of BH sub-systems in the merging scenario are in good agreement with the  observationally deduced SMBH masses. This is consistent with the observationally deduced dynamical $M/L_V$ ratios of these same SMBH-bearing UCDs which implies that the compact sub-cluster at the centre of the present-day UCDs can be observationally interpreted as a SMBH.

\item Moreover, we found that the same initial conditions describing well the observed  $M/L_{\rm V}$ ratios also reproduce the observed relation between the half-mass radii and the present-day masses, $M$, of UCDs.

\end{enumerate}

In view of the models computed here,  we conclude that the observed UCDs with large $M/L_V$ values most likely formed monolithically more than $5\,$Gyr ago with an IMF which becomes more top heavy with increasing mass and with half-mass radii consistent with those inferred as the birth radii for present-day embedded clusters observed to form in the disk of the Galaxy. This is broadly consistent with the model of UCD formation studied by \cite{Murray09} although the here-implied birth half-mass radii are significantly smaller. UCDs with small $M/L_V$ values  may have formed as merged complexes of embedded clusters (e.g. as are observed in the Antennae galaxies). Those in between would be a mixture of a monolithically formed component which merged with other star clusters in a centrally concentrated cluster complex. These can achieve the larger observed radii as shown by \cite{BruensKroupa11} but are, according to the present stellar population models, mostly those with the small $M/L_{\rm v}$ ratios at a given value of $M$. It is rather impressive how the data follow the  models calculated here (Figs.~\ref{fig:Z_RF_mass_dependent} and~\ref{fig:RM}), whereby the models are constructed using entirely independent observational constraints.

Our computations are based on the adopted initial half-mass-radius--mass relation (Eq. \ref{eq:rMrel}) derived by \cite{Marks12} which is extreme at the high mass end (a birth UCD mass of $10^8 M_{\odot}$ gives a half-mass radius of about 1 pc) given that young massive clusters obey the region called the “zone of avoidance” \citep{Noris_2014, Misgeld_2011}.
As an upcoming project, we plan to investigate how the results will change (given the dependence of the IMF on density) if a mass-radius relation similar to the observed size-mass relation of \cite{Dabringhausen08} which is similar to that derived by \cite{Murray09} is used instead of the here assumed \cite{Marks12} relation. This will allow us to test the UCD formation scenarios studied by \cite{Murray09}, as well as testing two possible extremes for the initial radius-mass relation.

\section*{Acknowledgements}

TJ acknowledges support through a two-year ESO Studentship in
Garching, the Erasmus+ programme of the European Union under
grant number 2017-1-CZ01- KA203-035562 on La Palma, and support through the ESA fellowship program. PK and TJ acknowledge support from the Grant Agency of the Czech Republic under grant number 20-21855S and from the DAAD-East European partnership programm 2020 at Bonn University.

\section*{Data Availability}
The data underlying this article are available in \citealt{Mieske08, Mieske13}



\bibliographystyle{mnras}
\bibliography{pap_UCDform.bbl}


\bsp	
\label{lastpage}
\end{document}